\documentclass[11pt]{article}  
\usepackage{ltexpprt}
\usepackage{sodafixes}

\usepackage{graphicx}
\usepackage{cite}
\usepackage{url}
\usepackage{listings}
\DeclareSymbolFont{AMSb}{U}{msb}{m}{n}
\DeclareSymbolFontAlphabet{\Bbb}{AMSb}
\def\R{\ensuremath{\Bbb R}}

\mathcode`O="724F

\begin{document}

\title{\Large Testing Bipartiteness of Geometric Intersection Graphs}
\author{David Eppstein\thanks{School of Information and Computer Science,
University of California, Irvine, CA 92697-3425, USA, eppstein@ics.uci.edu}}

\date{ }
\maketitle
\begin{abstract}
We show how to test whether an intersection graph of $n$ line segments or simple polygons in the plane, or of balls in $\R^d$, is bipartite, in time $O(n\log n)$.  More generally we find subquadratic algorithms for connectivity and bipartiteness testing of intersection graphs of a broad class of geometric objects. Our algorithms for these problems return either a bipartition of the input or an odd cycle in its intersection graph. We also consider lower bounds for connectivity and $k$-colorability problems of geometric intersection graphs.
For unit balls in $\R^d$, connectivity testing has equivalent randomized complexity to construction of Euclidean minimum spanning trees, and for line segments in the plane connectivity testing has the same lower bounds as Hopcroft's point-line incidence testing problem; therefore, for these problems, connectivity
is unlikely to be solved as efficiently as bipartiteness. For line segments or planar disks, testing $k$-colorability of intersection graphs for $k>2$ is NP-complete.
\end{abstract}

\pagestyle{plain}
\thispagestyle{empty}

\section{Introduction}

Suppose we are given a collection of geometric objects, for instance line segments in the plane.
Can we partition them into a small number of subsets, such that within each subset the objects are disjoint from each other?  This geometric problem can be modeled graph-theoretically as that of coloring an \emph{intersection graph}: if we form a graph that has a vertex per object and an edge connecting any two non-disjoint objects, the partition we seek is equivalent to a coloring of the graph, and the number of subsets in the partition equals the number of colors in the corresponding coloring.  For three or more colors, graph coloring is NP-complete, and remains so for many natural classes of intersection graphs, so we are unlikely to solve partition problems of this type efficiently.  But what if we wish to partition the input into only two non-self-intersecting subsets?  Testing bipartiteness for graphs can easily be solved in time linear in the size of the input graph, but this result does not necessarily translate into the most efficient algorithms for the geometric partitioning problem, as the corresponding graph bipartiteness instance may have quadratic size. Constructing the graph and then testing it for bipartiteness could take time quadratic in the number of input objects.  Can we do better?

\begin{figure*}[t]
\centering
\includegraphics[height=2.3in]{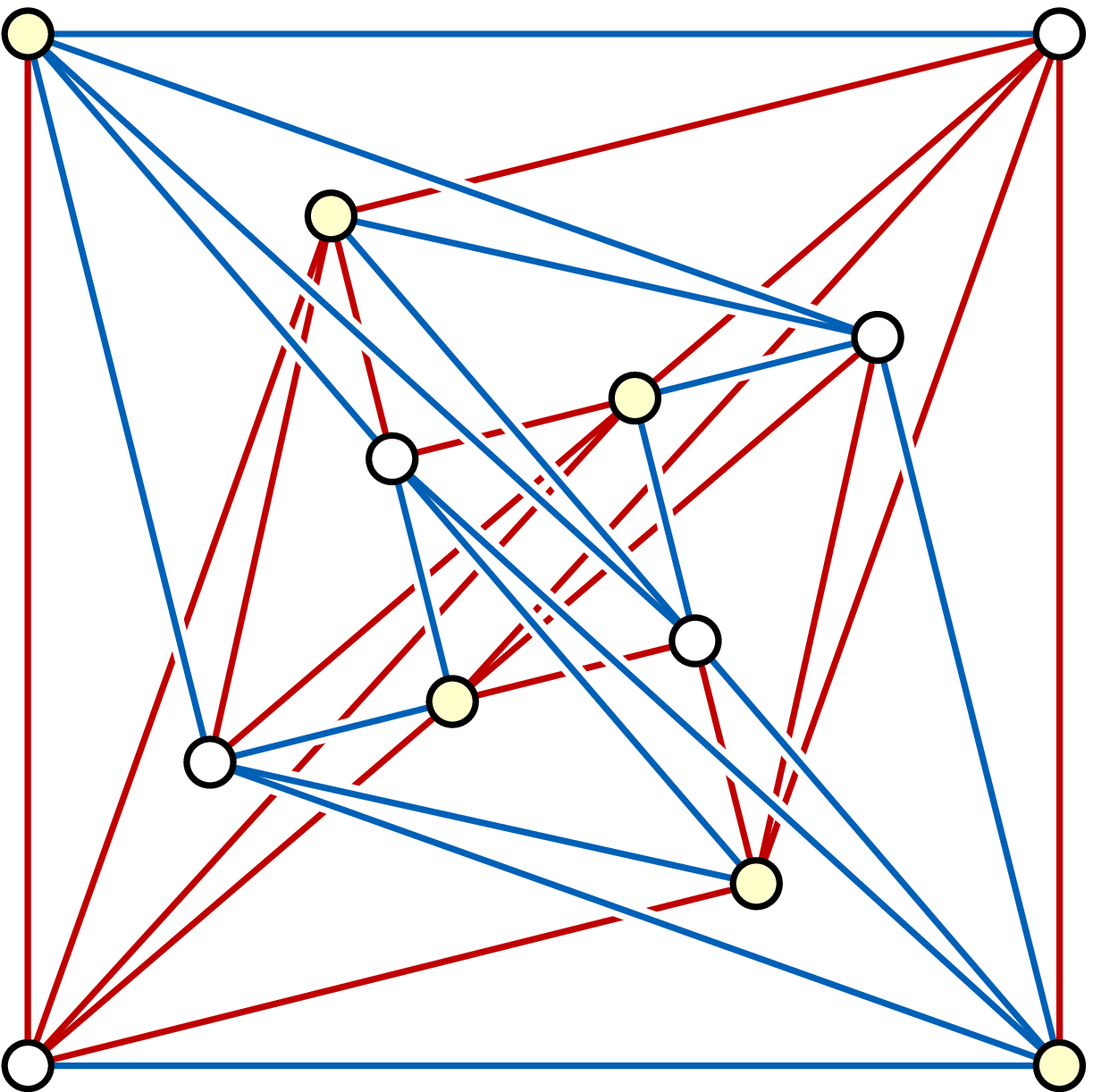}
\qquad\qquad
\includegraphics[height=2.7in]{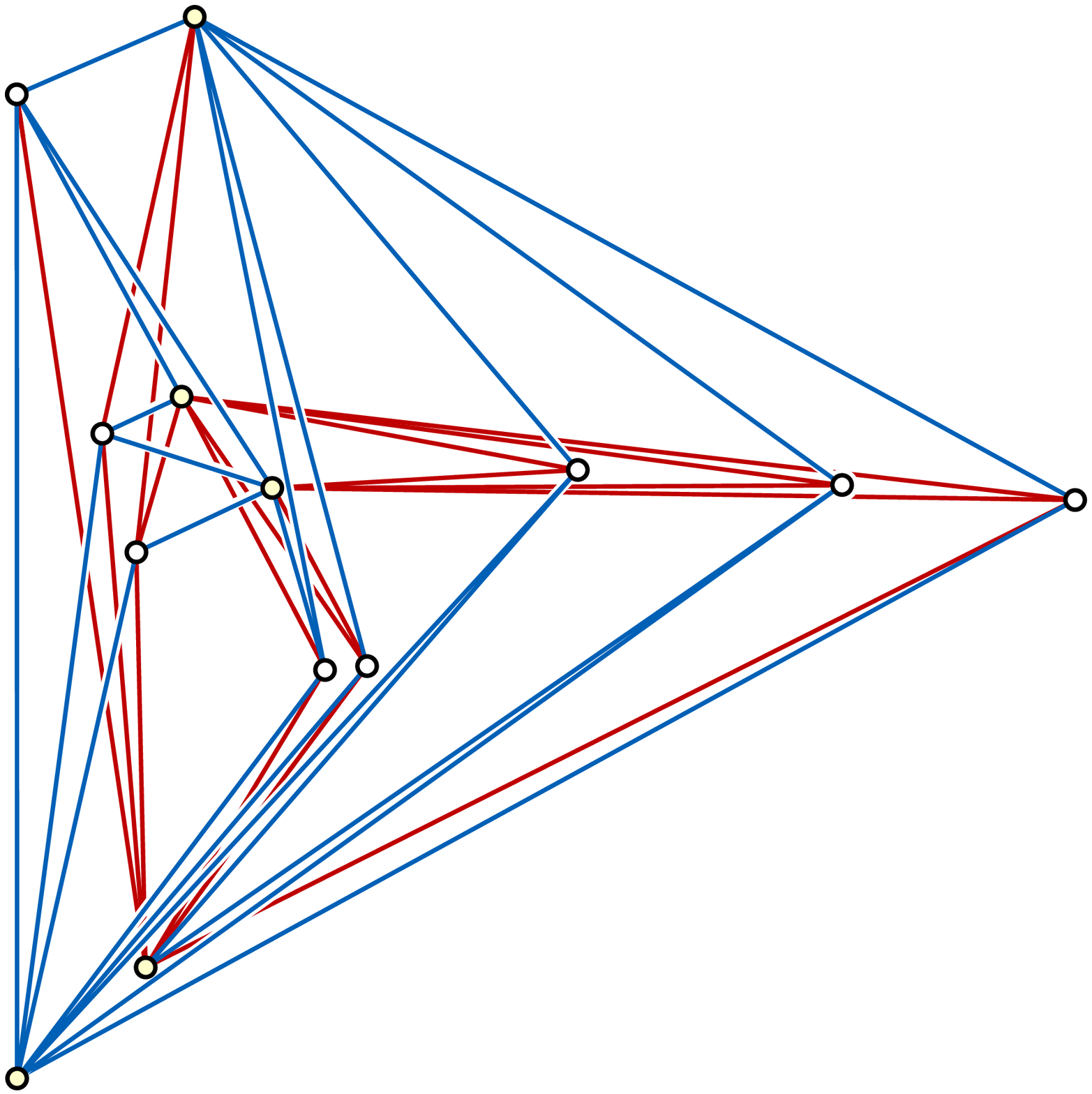}
\caption{Geometric thickness two drawing of $K_{6,6}$ (left) and $K_{5,8}$ (right).
The partitions of the drawings' edges into two non-crossing subsets
are indicated by the coloring and casing of the edges. The existence of such a partition is equivalent to the intersection graph of the segments being bipartite.}
\label{fig:K66}
\end{figure*}

The motivation for our study comes from graph drawing, specifically
\emph{geometric thickness}~\cite{BarMatWoo-EJC-06,DilEppHir-JGAA-00,DunEppKob-SCG-04,DujWoo-GD-05,Epp-GD-02,Woo-DM-03}, also known as \emph{real linear thickness}~\cite{Kai-AMSUH-73}, or, in the case of geometric thickness two,
\emph{doubly-linear graphs}~\cite{HutSheVin-GD-95}.  A \emph{drawing} of a graph, for our purposes, consists of an assignment of the graph's vertices to points in the plane, and its edges to simple plane curves connecting these points, such that no vertex lies on an edge unless it is an endpoint of that edge.  A \emph{straight-line drawing} is a drawing in which all the edges are drawn as line segments.  The \emph{thickness} of a drawing is the minimum number of subsets into which the graph's edges can be partitioned such that, within any subset, no two edges are drawn as curves that cross each other.  Alternatively, we can imagine assigning colors to the edges in a drawing, such that each color class forms a planar drawing.
A graph's thickness~\cite{Kai-AMSUH-73} is equal to the minimum thickness of any drawing of the given graph, while its geometric thickness is the minimum thickness of any straight-line drawing. The geometric thickness is at most the thickness, but may differ from the thickness by an arbitrarily large factor~\cite{Epp-TTGG-04,BarMatWoo-EJC-06}. Figure~\ref{fig:K66} shows thickness-two straight-line drawings of two complete bipartite graphs; these graphs are non-planar, so their geometric thickness is two.  The color separation between crossing edges in low-thickness drawings may reduce visual ambiguity when following connections between vertices.

We would like to include in the graph drawing repertoire procedures for finding low-thickness drawings of graphs.  However, testing the thickness of a graph is at most some given value $k$ is NP-complete, even for $k=2$~\cite{Man-MPCPS-83}.  We cannot yet prove a similar NP-completeness result for geometric thickness but it seems likely to be true as well.
Since exact computation of geometric thickness appears to have high computational complexity,
we are led to other procedures:
either interactive drawing processes in which a human operator adjusts vertex placements in an attempt to reduce the drawing's thickness, or heuristic search procedures for finding low-thickness placements.  In either case, in order to implement these procedures, we must determine the thickness of individual drawings of graphs, a simpler problem than finding a low-thickness drawing.  The problem of testing whether a given straight-line drawing has thickness two is exactly the bipartiteness problem for the line segment arrangement formed by the drawing's edges.

Another graph drawing application arises in the \emph{casing} of drawings with crossings; that is, the insertion of breaks into one of the edges at each crossing in order to create the visual appearance that the broken edge passes underneath the unbroken one.   Eppstein et al.~\cite{EppKreMum-WADS-07} define a \emph{switch} in a cased drawing to be a pair of adjacent crossings on a single edge $e$ such that $e$ is drawn as passing above its crossing edge in one of the crossings and below its crossing edge in the other crossing. They show that the problem of finding a casing that minimizes the number of switches can be solved in polynomial time, but the solution is complex as it involves reducing the problem to the Chinese Postman Problem. Our algorithm for testing bipartiteness of the intersection graph of the edges of the drawing allows us to test more efficiently whether a casing with no switches is possible, as such a casing exists if and only if the edges that participate in crossings can be partitioned into two subsets, one subset consisting of edges that only pass over their crossing edges and the other subset consisting of edges that only pass under their crossing edges.

Partitions into few non-crossing subsets may have other algorithmic uses as well.  For instance, vertical ray shooting in arrangements of non-crossing line segments can be performed in logarithmic time and linear space by point location in the trapezoidal decomposition of the segments, while achieving the same query time for arrangements of crossing line segments seems to require quadratic space.  If the arrangement can be partitioned into few non-crossing subsets, a vertical ray shooting query can be handled by combining independent query results within each subset, providing a significant space savings.

\begin{figure}[t]
\centering\includegraphics[width=4in]{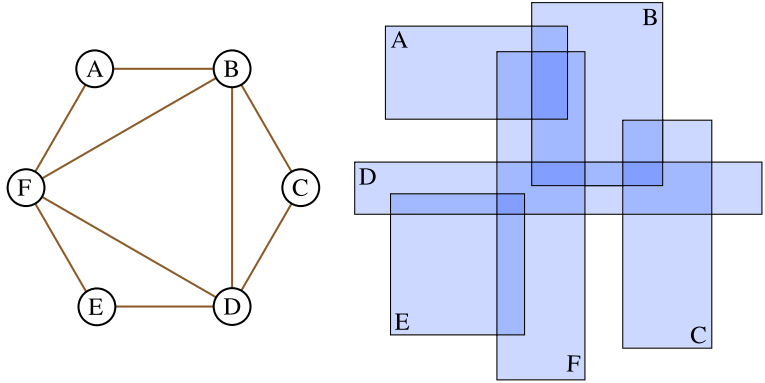}
\caption{An intersection graph of rectangles.}
\label{fig:intersection-graph}
\end{figure}
The problems we consider in this paper are of the following type.  Given a collection of geometric objects, such as line segments, their intersection graph is an undirected graph that has one vertex per object, and connects two objects by an edge whenever the intersection of the two objects is nonempty; an example intersection graph is depicted in Figure~\ref{fig:intersection-graph}.  We wish to determine whether the intersection graph is bipartite, and if so to find a two-coloring of this graph.  In the case of geometric thickness, the objects we are interested in are the edges of the drawing (that is, they form an arrangement of open line segments) and we wish to test whether the intersection graph of these line segments is bipartite.  Due to this graph drawing application our main interest in this paper and our main results concern intersection graphs of line segments, but it may be of interest in other applications to test bipartiteness of other intersection graphs such as those of disks or rectangles, or intersection graphs in higher dimensions, and we consider problems of these types as well.

Formally, for a class $\cal F$ of geometric objects (line segments, disks, rectangles, etc.), any finite set $X\subset\cal F$ defines an intersection graph $I(X)$ that has the objects of $X$ as vertices, with an edge between objects $x$ and $y$ whenever $x\cap y\ne\emptyset$. We
define the \emph{bipartiteness testing problem} for $\cal F$ to be the problem of determining whether $I(X)$ is bipartite for a given input $X\subset\cal F$, and we define the \emph{bipartition problem} for $\cal F$ to be the problem of either finding a bipartition of $I(X)$, if one exists, or reporting that no such bipartition exists. If no bipartition exists, we may additionally desire that our algorithms return an odd cycle in the intersection graph, so that users of the algorithm may easily verify its correctness.

For many types of objects, including line segments, disks, and rectangles, algorithms are known that can construct the intersection graph in time that matches or nearly matches the output size~\cite{ChaEde-JACM-92,EppMilTen-FI-95}. One could then apply any linear-time graph bipartiteness algorithm to the intersection graph to solve either the bipartiteness testing or bipartition problem.  However, in the worst case these graphs can have quadratic size; for instance in the graph thickness problem, it may be common to have many nearly-horizontal segments in one layer and many nearly-vertical segments in another layer, creating a large number of crossings. Thus, in the worst case, these algorithms would have running time $\Theta(n^2)$. Our goal in this paper is to improve on this naive approach, and achieve subquadratic or even near-linear complexity bounds for the intersection graph bipartiteness testing problem and bipartition problem. The bounds we achieve will vary, depending on the types of geometric object belonging to $\cal F$.

Curiously, while for graphs connectivity is somewhat simpler than bipartiteness,
for geometric intersection graphs the situation seems to be reversed:
only one of our bipartiteness testing algorithms extends directly to connectivity.
We provide evidence that, for unit balls in dimensions greater than two, bipartiteness testing is strictly easier than connectivity, by showing an equivalence between connectivity testing and Euclidean minimum spanning trees: our time bounds for bipartiteness testing for balls are significantly faster than the best known Euclidean minimum spanning tree algorithms in these dimensions. For planar line segments, also, our bipartiteness testing algorithm has time bounds significantly smaller than lower bounds on connectivity testing derived by a reduction from Hopcroft's problem on point-line incidences~\cite{Eri-DCG-96}.

\section{New Results}

We provide the following new results:

\begin{figure}[t]
\centering\includegraphics[width=4in]{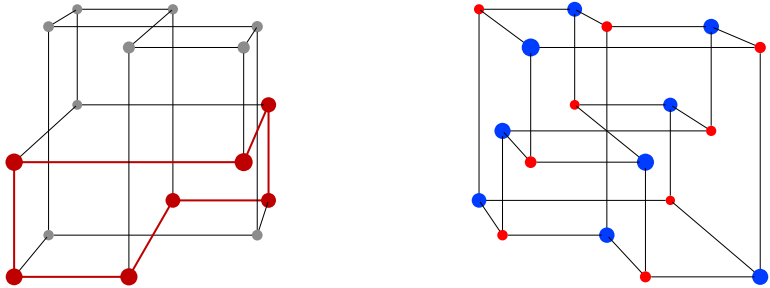}
\caption{Witnesses for bipartiteness or nonbipartiteness, of the type returned by our algorithms: an odd cycle in a nonbipartite graph (left) or a bipartition of a bipartite graph (right).}
\label{fig:witnesses}
\end{figure}

\begin{itemize}
\item We describe a generic technique that applies to many natural classes of geometric objects with bounded description complexity, such as polygons with a bounded number of sides or algebraic curves with bounded degree. For these classes, this technique leads to subquadratic times for testing the bipartiteness of the objects' intersection graphs and finding a bipartition of these graphs. The same technique may also be used for listing the connected components of the objects' intersection graphs.

\item We provide an $O(n\log n)$ time algorithm for testing bipartiteness of $n$ balls in $\R^d$, based on a separator theorem for sets of balls with no triple intersection and a separator-based divide and conquer technique. We also
use a randomized reduction to show that testing connectivity of unit balls is equivalent in complexity to constructing Euclidean minimum spanning trees, and therefore is unlikely to have near-linear algorithms in dimensions higher than two.

\item We give an $O(n\log n)$ time algorithm for testing bipartiteness of $n$ line segments in the plane, or of a collection of simple polygons with $n$ total edges.  Our technique is based on a plane sweep algorithm, together with new data structures for maintaining information about the possible bipartitions of the part of the arrangement that has been swept by the algorithm. However, this technique does not extend to connectivity: as we observe, finding the connected components of line segments is at least as hard as Hopcroft's problem of finding point-line incidences and is therefore unlikely to have as efficient an algorithm.

\item We construct reductions proving the NP-completeness of testing the 3-colorability of intersection graphs for line segments or unit disks in the plane.
\end{itemize}

All our bipartiteness testing algorithms provide a \emph{witness} for their result: they return either a coloring of the input objects, in the case that the input is bipartite, or an odd cycle in the intersection graph if it is not (Figure~\ref{fig:witnesses}).

There has been some prior work on connectivity of geometric intersection
graphs: Imai~\cite{Ima-IPL-82} describes an algorithm for partitioning a set of squares in the plane into connected components, and Imai and Asano~\cite{ImaAsa-Algs-83} extend this method to rectangles.
Guibas et al.~\cite{GuiHerSur-DCG-01} describe data structures for maintaining the connected components of disks in the plane, subject to both continuous motion of the disks and discrete changes in the set of disks or in their motions, and Hershberger and Suri~\cite{HerSur-SCG-99,HerSur-SODA-01} study similar problems for rectangles. However, we are unaware of prior work on intersection graph bipartiteness.

\section{Generic Bipartiteness Algorithm}

In this section we describe a generic approach to intersection graph bipartiteness testing, that works for a wide class of geometric objects including polygons in the plane with a bounded number of sides, algebraic curves with bounded degree, or simplices in any bounded dimension. The only requirement for applying this approach is that there exist a data structure that allows us to find the objects that intersect a query object, among some set of objects, and to delete objects from the set. Typically, when this approach works, it provides algorithms with subquadratic running time, improving the $O(n^2)$ time bound of an algorithm that explicitly constructs and then tests the intersection graph itself. However, the general purpose nature of this algorithm means that, as we will see in subsequent sections of this paper, it can often be improved for more specific classes of geometric object.

Before we describe the algorithm in detail, let us broadly outline our approach. Consider standard graph-theoretic bipartiteness testing algorithms: for instance, one typical approach is to perform a depth first search of the input graph, partitioning the graph into \emph{tree edges} that form part of the depth first search tree and \emph{back edges} that connect any vertex with one of its ancestors in the depth first search tree. The tree can be partitioned into two color classes by placing vertices an even distance from the root into one class and vertices an odd distance from the root into the other class; the graph is bipartite if and only if the resulting partition is a proper coloring, which may be tested by examining each edge of the graph to determine whether its end points have different colors. If we wish to make this algorithm more efficient in the geometric setting, we must concentrate our attention on two of its steps. First, in performing the depth-first search itself, we will not have time to examine all neighbors of a given intersection graph vertex; instead, we need to quickly locate neighbors that are not already part of the depth first search tree, as only those neighbors need to be explored. And second, when testing whether the edges incident to each vertex are properly colored, we do not have time to examine all such edges; instead, we need to quickly identify whether there is an edge that connects the vertex to another vertex within the same color class. Both of these tasks involve maintaining sets of geometric objects (corresponding to the vertices that have not yet been explored, or the vertices within a single color class, respectively) and testing whether a query object intersects anything in the set.

To formalize the type of data structure needed for this task, we define a class of data structures which we call \emph{intersection detection} data structures.
An \emph{intersection detection} data structure for a set $S$ of objects is one that can handle \emph{queries} of the following type: we are given as an argument to the query an object $x$ of the same type as the ones in $S$, and must return as the result of the query an object in $S$ that intersects $x$, if one exists.  An intersection detection data structure is \emph{decremental} if it also supports \emph{deletion} operations that remove a specified object from $S$, so that subsequent queries will not return the deleted object.  We measure the time complexity of a decremental intersection detection data structure $\cal D$ by two functions:
$Q_{\cal D}(n)$, measuring the amortized time per query when the size of the set of objects is $n$,
and $T_{\cal D}(n)$, measuring the time to set up a data structure for and then delete in worst-case order all the objects in a set of $n$ objects.   For most natural classes of geometric object with bounded description complexity, standard range searching techniques~\cite{AgaEri-ADCG-99}
can be used to provide decremental intersection detection data structures, where
$T_{\cal D}(n)=O(n^{2-c})$ and $Q_{\cal D}(n)=O(n^{1-c})$ for some constant $c>0$ that depends only on the object type.

\begin{lemma}
\label{lem:did-sf}
Let $\cal C$ be a class of objects having a decremental intersection detection data structure ${\cal D}$ with $T_{\cal D}(n)=O(n^{2-c})$ and $Q_{\cal D}(n)=O(n^{1-c})$.
Then in time $O(n^{2-c})$ we can construct a spanning forest for the intersection graph
of any set of $n$ objects in class~$\cal C$.
\end{lemma}

\begin{proof}
We perform a depth first search of the intersection graph, and return the resulting depth first spanning forest.  At each step of the depth first search we maintain a decremental intersection detection data structure representing the objects that have not yet been visited in the search; initially this consists of all the objects in the input.  When the search reaches a previously unvisited object~$x$, we add $x$ to the spanning forest, remove $x$ from the intersection detection data structure, repeatedly use the intersection detection data structure to find unvisited objects intersecting~$x$, and recursively visit each such object.  Each query either leads to a new visited object or is the last query for $x$, so the algorithm makes fewer than $2n$ total queries
and the running time is as stated.
\end{proof}

\begin{theorem}
\label{thm:generic}
Let $\cal C$ be a class of objects having a decremental intersection detection data structure ${\cal D}$ with $T_{\cal D}(n)=O(n^{2-c})$ and $Q_{\cal D}(n)=O(n^{1-c})$.
Then in time $O(n^{2-c})$ we can test the bipartiteness of an intersection graph of any set of $n$ objects in class~$\cal C$, and return either a two-coloring of the intersection graph or an odd cycle in the intersection graph.
\end{theorem}

\begin{proof}
We first apply the spanning forest algorithm of Lemma~\ref{lem:did-sf} to construct a spanning forest $F$ of the set of input objects. We use $F$ to two-color the objects,
using one color for objects at even height in the forest and the other color for objects at odd height.  
To test the validity of the resulting coloring, we search for intersections within each color class, by applying the spanning forest algorithm two more times, once each for the sets of objects in the two color classes: if either color class has a spanning forest with a nontrivial tree, then we have an intersection between two objects of the same color. If no two objects of the same color intersect, we return the coloring.  If two objects $x$ and $y$ having the same color intersect, then $x$ and $y$ must be within the same tree of $F$, because the trees of $F$ coincide with the connected components of the intersection graph; in this case, the intersection graph is nonbipartite, as an odd cycle in the intersection graph can be found by adding edge $xy$ to the path in $F$ that connects $x$ to $y$.
\end{proof}

We remark that a decremental intersection detecting data structure may often be constructed from a static structure that allows intersection detection queries but not deletions, by partitioning the dynamic set of items into many small subsets and maintaining a static subset for each item.
As an example of this approach, consider the problem of testing the bipartiteness of a set of lines in $\R^3$. Chazelle~\cite{Cha-Algo-96} describes data structures for determining the orientation of a single line with respect to $n$ other lines, which can be used to perform static intersection detection in $O(\log n)$ time per query after $O(n^{4+\epsilon})$ preprocessing time, for any $\epsilon>0$. If we partition the $n$ lines into $n^{4/5}$ subsets of $n^{1/5}$ lines each, and maintain a data structure of this type for each subset, we may perform intersection detection queries by testing each subset separately in time $O(n^{4/5}\log n)$ per query, and handle any deletion by rebuilding the data structure for a single subset in time $O(n^{4/5+\epsilon})$. Thus, by Theorem~\ref{thm:generic}, we may determine the bipartiteness of the intersection graph of a set of $n$ lines in $\R^3$, in time $O(n^{9/5+\epsilon})$, slightly improving the much simpler $O(n^2)$ time bound achieved by computing the intersection graph of the lines.

For this example of lines in space, and for most of the other bipartitness testing problems we consider, this generic bipartiteness testing approach will lead to relatively slow and complex algorithms, due to the difficulty of the decremental intersection detection problem. Therefore
we view Theorem~\ref{thm:generic} not as our main result but rather as a baseline for further algorithmic improvements.  As we shall see in subsequent sections, the time bounds coming from this theorem can be significantly improved in certain important cases, including bipartiteness testing for intersection graphs of balls in $\R^d$ and of line segments in the plane.

\section{Bipartiteness for Balls}

We now describe an alternative approach for testing whether the intersection graph of a set of balls in $\R^d$ is bipartite, in $O(n\log n)$ time for any fixed dimension~$d$.  We then provide a randomized reduction showing the equivalence (to within polylogarithmic factors) of the problems of computing Euclidean minimum spanning trees and of constructing connected components of unions of unit balls; this equivalence, together with known results on spanning tree construction, shows that it is unlikely that we can construct connected components as quickly as we can test bipartiteness.

The basic idea of our bipartiteness testing algorithm is that, while ball intersection graphs in general can be dense, bipartite ball intersection graphs must be sparse.  More generally, similar sparsity results are true for $k$-ply ball arrangements~\cite{MilTenThu-JACM-97}, that is, arrangements of balls in which each point of space is covered by at most $k$ balls; a bipartite intersection graph of balls is a $k$-ply arrangement with $k=2$.  Our proof follows an approach used by Teng~\cite{Ten-PhD-91} for a related but different family of geometric graphs, the \emph{overlap graphs} formed by connecting two balls by an edge when a constant dilation of the smaller ball intersects the larger one.

\begin{figure*}[t]
\centering
\includegraphics[height=2in]{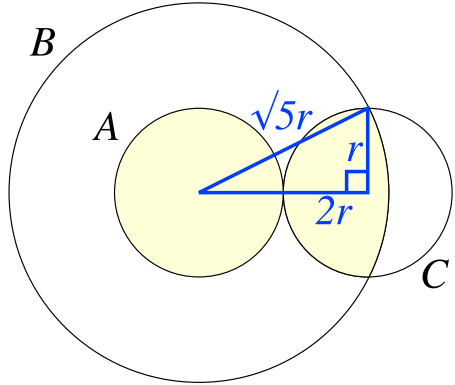}
\caption{Illustration for Lemma~\ref{lem:sqrt5-tan}. The intersection between one of two tangent balls and the $\sqrt 5$ dilation of the other ball has volume at least half that of the original balls.}
\label{fig:sqrt-tan}
\end{figure*}

\begin{lemma}
\label{lem:sqrt5-tan}
Let $A$, $B$, and $C$ be three balls in $\R^d$ such that $A$ and $B$ are concentric, $A$ and $C$ are tangent, $A$ and $C$ both have radius $r$, and $B$ has radius $\sqrt 5 r$. Then the volume of $B\cap C$ is at least half the volume of $A$.
\end{lemma}

\begin{proof}
Let $p$ be the point of tangency, $q$ be the point of $C$ farthest from $p$, and $E$ be the ``equator'' of $C$, the points on its boundary that are equidistant from $p$ and $q$. Let $a$ and $c$ be the centers of $A$ and $C$ respectively, and let $b$ be any point of $E$. Then, $|ac|=2r$, $|bc|=r$, and line segments $ac$ and $bc$ are perpendicular, so we may apply the Pythagorean theorem to calculate that $|ab|=\sqrt 5 r$, as shown in Figure~\ref{fig:sqrt-tan}. That is, $B$, which has radius $\sqrt 5 r$ and is centered at $a$, contains all points of $E$, and therefore also contains all points of $C$ that are closer to $p$ than to $q$. These points form a volume of half that of $C$, but since $C$ and $A$ have the same volume they also form a volume of half that of $A$.
\end{proof}

\begin{figure*}[t]
\centering
\includegraphics[height=2in]{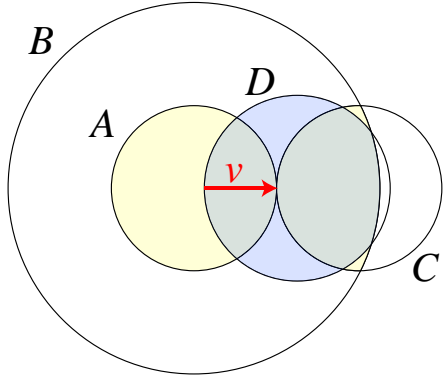}
\caption{Illustration for Lemma~\ref{lem:sqrt5-bigger}. The intersection between the larger of two intersecting balls and the $\sqrt 5$ dilation of the smaller ball has volume at least half that of the smaller ball.}
\label{fig:sqrt-bigger}
\end{figure*}

\begin{lemma}
\label{lem:sqrt5-bigger}
Let $A$, $B$, and $D$ be three balls in $\R^d$ such that $A$ and $B$ are concentric, $A$ and $D$ intersect, $A$ has radius $r$, $D$ has radius at least as large as $r$, and $B$ has radius $\sqrt 5 r$. Then the volume of $B\cap D$ is at least half the volume of $A$.
\end{lemma}

\begin{proof}
Let $v$ be a translation vector parallel to the vector from the center of $A$ to the center of $D$, such that $D+v$ is tangent to $A$, and let $C$ be a circle with radius equal to $A$, tangent to $A$ at the point of tangency of $D+v$, as shown in Figure~\ref{fig:sqrt-bigger}.
Then $D+v\supset C$, so $(B\cap D)+v\supset B\cap C$. Therefore, the volume of $B\cap D$ is at least as large as the volume of $B\cap C$, which by Lemma~\ref{lem:sqrt5-tan} is at least half the volume of~$A$.
\end{proof}

\begin{lemma}
\label{lem:sqrt5-nbhd}
Let $A$ be a ball in $\R^d$, and let $S$ be a collection of disjoint balls, each larger than $A$ and each intersecting $A$. Then $|S|\le 2\cdot 5^{d/2}$.
\end{lemma}

\begin{proof}
Let $B$ be a ball concentric with $A$, with radius $\sqrt 5$ times that of $A$., and let $V$ denote the volume of $A$; then $B$ has volume $2V\cdot 5^{d/2}$. By Lemma~\ref{lem:sqrt5-bigger}, each ball in $S$ intersects $B$ in a set with volume at least $V/2$. Since these sets are disjoint, the total volume of $(\bigcup S)\cap B$ must be at least $|S|\cdot V/2$. But since $(\bigcup S)\cap B$ is a subset of $B$, it can have volume no more than $B$ itself; that is, $|S|\cdot V/2\le V\cdot 5^{d/2}$. The result follows by dividing both sides of this inequality by $V/2$.
\end{proof}

\begin{lemma}
\label{lem:sqrt5}
Let $B$ be a collection of $n$ balls in $\R^d$ having a bipartite intersection graph $G_B$.
Then $|E(G_B)|\le c_d n$ for some constant $c_d\le 2\cdot 5^{d/2}$.
\end{lemma}

\begin{proof}
Orient the edges of the intersection graph from smaller balls to larger ones, choosing the orientation arbitrarily in cases where both balls have equal radius. The neighbors of any ball in the intersection graph must be disjoint, else the intersection graph would contain a triangle, violating bipartiteness.
Therefore, by Lemma~\ref{lem:sqrt5-nbhd}, each ball has at most $2\cdot 5^{d/2}$ outgoing neighbors. Summing the number of outgoing neighbors of all vertices, we find that the graph has at most $2n\cdot 5^{d/2}$ edges, as was to be proven.\end{proof}

A similar technique of bounding the number of neighboring larger balls by adding up the volumes of their intersections with a dilated copy of the smallest ball is used by Teng~\cite{Ten-PhD-91}, e.g. in his Theorem 4.3. However, Teng uses a larger dilation factor: translated to our terms, his proof technique would expand each ball by a factor of three rather than by $\sqrt 5$. Our more careful choice of dilation factor leads to a better dependence of $c_d$ on $d$.
The same proof technique, with the $\sqrt 5$ dilation factor, leads to a bound of $5^{d/2}2kn$ for the number of intersections in a $k$-ply ball arrangement, slightly improving a $3^d kn$ bound of Miller et al.~\cite[Theorem 3.3.1]{MilTenThu-JACM-97}.

The other ingredient we need is an algorithm for constructing intersection graphs of $k$-ply ball arrangements. To do this, we use a divide-and-conquer technique based on the following separator theorem.

\begin{lemma}[Eppstein et al.~\cite{EppMilTen-FI-95}]
\label{lem:ball-ig-separator}
Let there be given $n$ balls in $\R^d$ with at most $k$ balls intersecting at any point, and let $\epsilon>0$ be any constant.
Then in time $O(n)$ we can construct a sphere $S$ such that $S$ intersects $O(k^{1/d}n^{1-1/d})$ of the balls, where neither the interior nor the exterior of $S$ contains more than $(1+\epsilon)(d+1)n/(d+2)$ balls.
\end{lemma}

\begin{figure}[t]
\centering
\includegraphics[height=3in]{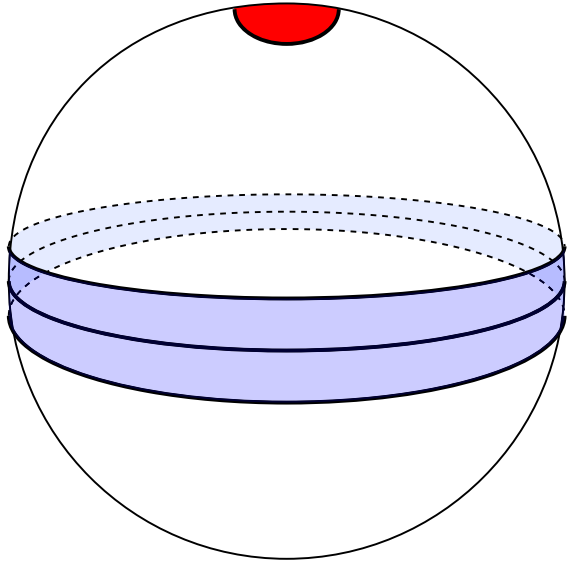}
\caption{A disk on a sphere (red) and the corresponding equatorial belt (pale blue).}
\label{fig:equatorial-belt}
\end{figure}

We refer readers to Eppstein et al.~\cite{EppMilTen-FI-95} for the detailed proof of this lemma, which is beyond the scope of this paper. However, we describe here a sketch of the sphere separator construction algorithm used to prove this result. The algorithm performs the following steps:
\begin{itemize}
\item It performs a polar projection of the input balls onto a unit sphere in $\R^{d+1}$, so that each ball is transformed into a disk on the surface of the sphere.
\item It computes a \emph{centerpoint} $p$ of the centers of the transformed balls. That is, $p$ is a point in $\R^{d+1}$ (necessarily interior to the unit sphere) such that any hyperplane through $p$ contains at most $(d+1)n/(d+2)$ ball centers, and therefore will completely contain at most $(d+1)n/(d+2)$ balls. The existence of a centerpoint follows as a consequence of Helly's theorem, and it can be constructed as any point within the intersection of a certain family of $O(n^{d+1})$ halfspaces; however, this construction algorithm would not fit within the stated time bounds. Instead, efficient approximation algorithms for the centerpoint construction problem are known~\cite{ClaEppMil-IJCGA-96} that allow one to find an approximate centerpoint, such that any hyperplane contains at most $(1+\epsilon)(d+1)n/(d+2)$ balls, in time $O(n)$.
\item It performs a projective transformation of $\R^{d+1}$ to itself that maps the unit sphere to itself but transforms the computed centerpoint $p$ to the origin of $\R^{d+1}$ at the center of the sphere. This transformation preserves hyperplanes in $\R^{d+1}$, and since a ball on the unit sphere is formed by the intersection of the unit sphere with a hyperplane it also maps the original set of balls to a transformed set of balls, preserving the incidence properties of these balls with other hyperplanes. After this transformation, any hyperplane through the origin will contain at most $(1+\epsilon)(d+1)n/(d+2)$ of the transformed balls.
\item For each ball $b$, centered at a point $o_b$ with radius $r_b$, it forms a \emph{great hypercircle} by intersecting the unit sphere with a hyperplane through the origin perpendicular to the vector from the origin to $o_b$. It then replaces the ball by an \emph{equatorial belt} consisting of the points within distance $r$ from this great hypercircle (Figure~\ref{fig:equatorial-belt}). This replaced system of objects has the property that a hyperplane through the origin intersects one of the balls if and only if the line perpendicular to the hyperplane through the origin intersects the corresponding equatorial belt. It can be shown via H\"older's inequality (e.g., see~\cite{MilTenTHu-SJSC-98}, Theorem 6.1 and~\cite{Ten-PhD-91}, Lemma 10.5) that the sum of the volumes of the equatorial belts is $O(k^{1/d}n^{1-1/d})$; this bound therefore also gives the expected number of belts that contain a randomly chosen point on the sphere.
\item The algorithm applies an $1/r$-cutting-based prune and search algorithm to find point $q$ on the unit sphere covered by a number of belts that is at most the average covering number. That is, $q$ belongs to $O(k^{1/d}n^{1-1/d})$ great belts. The prune and search algorithm finds an $1/r$-cutting of the boundaries of the belts; that is, a triangulation of the unit sphere such that each simplex of the triangulation is crossed by at most a $1/r$ fraction of the belts, for some constant $r>0$. It then computes the average covering number within each simplex, and continues recursively within the simplex with smallest average covering number. In any fixed dimension, the time to construct the $\epsilon$-cutting and compute the average covering numbers within each of its simplices is linear; the total time for this step adds in a geometric series to $O(n)$.
\item The desired separator is formed by cutting the system of balls by a hyperplane perpendicular to the line from the origin through $q$. Reversing the 
system of transformations performed at the start of the algorithm, the intersection of this hyperplane with the unit sphere in $\R^{n+1}$ becomes a sphere $S$ in $\R^n$. The $O(k^{1/d}n^{1-1/d})$ balls cut by $S$ are included in the separator, and there are at most $(1+\epsilon)(d+1)n/(d+2)$ balls inside of it and at most $(1+\epsilon)(d+1)n/(d+2)$ balls outside of it.
\end{itemize}

We observe that all of these steps may be performed equally efficiently for any arrangement of balls, not necessarily satisfying the intersection condition: the time analysis of the algorithm doesn't depend on the assumption that the input system of balls has the $k$-ply property. However, if the input is not $k$-ply, the resulting sphere $S$ may intersect a larger number of balls than the stated $O(k^{1/d}n^{1-1/d})$ bound.

By combining Lemma~\ref{lem:ball-ig-separator} with a simple separator-based divide and conquer technique, the authors of that paper also proved the following:

\begin{lemma}[Eppstein et al.~\cite{EppMilTen-FI-95}]
\label{ball-ig-construction}
For any fixed $d$, the intersection graph of a system of $n$ balls in $R^d$, having at most $k$ balls intersecting at any point, can be constructed in time $O(kn+n\log n)$.
\end{lemma}

The algorithm achieving this time bound merely constructs a separator according to Lemma~\ref{lem:ball-ig-separator} and calls itself recursively on two subsets of balls: the set of balls inside the separating sphere together with the balls of the separator, and the set of balls outside the separating sphere together with the balls of the separator. The recursion stops when there are a constant number of balls, at which point the intersection graph may be constructed by testing all pairs of balls. Unlike the separator algorithm itself, the algorithm of Lemma~\ref{ball-ig-construction} does have a time complexity depending on the ply of the input system, since it produces an output graph that may be larger when larger numbers of intersections between balls are allowed.

As we now show, we can adapt this algorithm to test bipartiteness of ball intersection graphs efficiently. The key idea is that a bipartite intersection graph can cover each point of $\R^d$ with at most two balls, for otherwise the three or more balls covering some point would form a clique and contradict bipartiteness. Thus, when the intersection graph is bipartite, Lemma~\ref{ball-ig-construction} could be used to construct the graph and then 2-color it. However, our bipartiteness testing algorithm must still perform efficiently even when given non-bipartite inputs.

\begin{theorem}
For any fixed $d$, we can test the bipartiteness of an intersection graph of balls, and return either a two-coloring of the intersection graph or an odd cycle, in time $O(n\log n)$.
\end{theorem}

\begin{proof}
We apply a modified version of the algorithm of Lemma~\ref{ball-ig-construction}
to construct the intersection graph of the balls, and then apply a standard graph algorithm to find either a bipartition or an odd cycle in the constructed intersection graph. The modification to the algorithm consists in checking the size of each separator produced at each stage of the algorithm.  If, at each stage, the separator size is within the bounds expected for a system of balls covering each point in space at most twice, the algorithm will produce an intersection graph of size $O(n)$ which we can safely test for bipartiteness using a linear time graph algorithm.

If, however, some stage produces a separator that is too large, we terminate the intersection graph construction algorithm early without constructing the whole graph.  Recalling the proof of Lemma~\ref{lem:ball-ig-separator} sketched above, the separator can be large only if the average covering number of the system of equatorial belts constructed by the separator algorithm is too large, and this covering number can be large only if the system of transformed balls on the unit sphere in $\R^{d+1}$ has total volume more than twice the volume of the sphere.  That is, the average number of times each point on the unit sphere is covered by balls is more than two.  By using an $\epsilon$-cutting-based prune and search algorithm essentially identical to that of the algorithm of Lemma~\ref{lem:ball-ig-separator}, but applied to the transformed balls on the unit sphere instead of to their dual equatorial belts, we may in deterministic linear time find a point on the unit sphere covered by a number of balls at least equal to this average covering number. That is, more than two transformed balls cover this point on the unit sphere in $\R^{d+1}$, and by reversing the transformation that took our system of balls from $\R^d$ to a sphere in $\R^{d+1}$ we may find a point in $\R^d$ covered by more than two balls of our input arrangement. We then determine which balls cover this point and return as our odd cycle a triangle formed by three of these balls.
\end{proof}

It seems likely that a similar approach would work for more general classes of fat objects in
$\R^d$, such as polygons that have inscribed and circumscribed spheres of radii within a constant factor of each other. The main tool required is an appropriate separator theorem for those objects; some separator theorems for fat objects are known~\cite{SmiWor-FOCS-98}, but are less well developed than separator theorems for balls. We note that this $O(n\log n)$ time bound significantly improves the bound of the generic algorithm for connectivity or bipartiteness.

\begin{figure}[t]
\centering\includegraphics[height=2.5in]{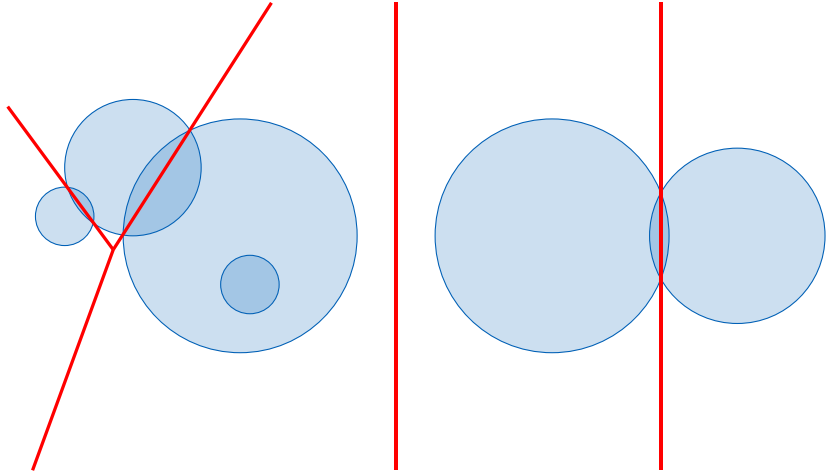}
\caption{A family of circles (blue) and a portion of their power diagram (red).}
\label{fig:powerdiag}
\end{figure}

We report as a curiosity the following structural result for bipartite intersection graphs of disks.

\begin{figure}[t]
\centering\includegraphics[width=4in]{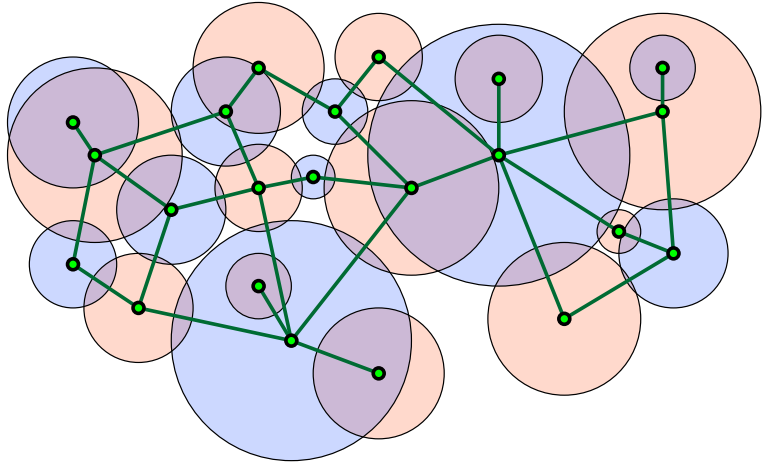}
\caption{Theorem~\ref{thm:disk-planarity}: a bipartite intersection graph of disks has a planar straight line embedding with the vertices at disk centers.}
\label{fig:disk-planarity}
\end{figure}

\begin{figure}[t]
\centering\includegraphics[width=3in]{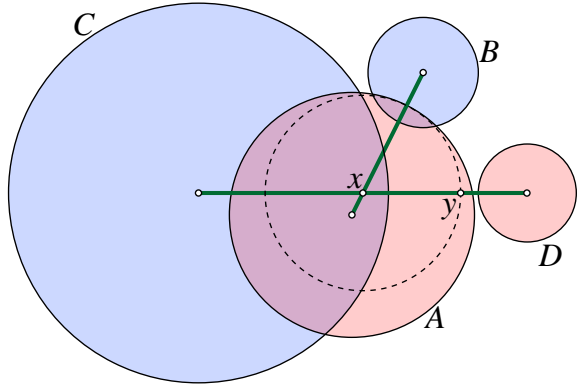}
\caption{Construction for proof of Theorem~\ref{thm:disk-planarity}: $A$ and $B$ are intersecting disks, $C$ and $D$ are any two other disks of opposite colors, $x$ is the point where the segments connecting the disk centers meet, and $y$ and the boundary of $A$ are equidistant from $x$.}
\label{fig:planarity-proof}
\end{figure}

\begin{theorem}
\label{thm:disk-planarity}
Let $B$ be a bipartite intersection graph of a set of disks in the plane. Then $B$ is a planar graph.
\end{theorem}

\begin{proof}
Assume without loss of generality (by perturbing the disks, if necessary) that no two disks are concentric, plase each vertex of the intersection graph at the corresponding disk center, and draw the edges of the intersection graph as straight line segments, as shown in Figure~\ref{fig:disk-planarity}. We claim that the resulting drawing has no crossings.

To see this, let $A$, $B$, $C$, and $D$ be any four disks of the input, such that $A$ and $B$ intersect, such that the line segments between the centers of $A$ and $B$, and between the centers of $C$ and $D$, cross at a point $x$, and such that the labeling of the disks is chosen in such a way that either $C$ and $D$ do not cross or $x$ is farther from the boundary of $A$ than it is from the boundary of $C$ and $D$. Further, choose the labeling in such a way that $C$ has a different color than $A$ and $D$ has the same color than $A$, let $d$ be the distance of $x$ from the boundary of $A$, and let $y$ be the point at distance $d$ from $x$ on the line segment from $x$ to the center of $D$ (Figure~\ref{fig:planarity-proof}). We claim that $C$ and $D$ do not cross, and therefore that this graph drawing can have no crossing. For, if $C$ and $D$ were to cross, then $x$ would be at distance at most $d$ from the boundary of $C$ by our choice of labeling, but it must be outside $D$ and farther than distance $d$ from $D$, for otherwise point $y$ would be a common intersection point of $A$ and $D$, violating bipartiteness. Thus, the line through $y$ perpendicular to the segment between the centers of $C$ and $D$ separates $C$ and $D$, so they cannot cross. Since no four disks $A$, $B$, $C$, and $D$ can form a crossing, the graph is planar.
\end{proof}

\begin{figure}[t]
\centering\includegraphics[width=3in]{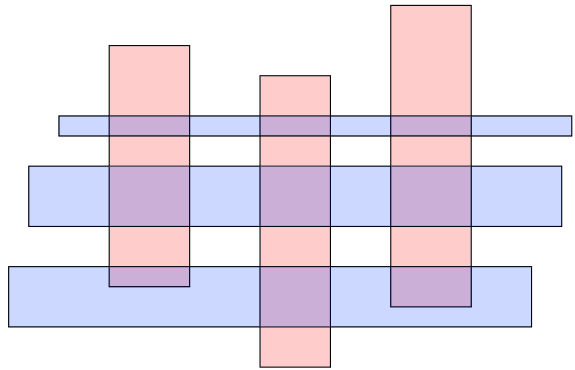}
\caption{$K_{3,3}$ as a rectangle intersection graph.}
\label{fig:k33-rectint}
\end{figure}

Thus, in the two-dimensional case, we can greatly strengthen Lemma~\ref{lem:sqrt5}: that lemma shows a $10n$ bound on the number of edges of a bipartite intersection graph of disks, but every planar bipartite graph has at most $2n-4$ edges.
Bipartite intersection graphs of other planar objects need not be planar; for instance, the nonplanar graph $K_{3,3}$ can be realized as a rectangle intersection graph (Figure~\ref{fig:k33-rectint}).

It is possible to construct a spanning forest for the connected components of a set of disks in the plane, regardless of whether the intersection graph is bipartite, in the same $O(n\log n)$ time bound, as we describe below. The \emph{power} of a point $p$ with respect to a disk centered at point $q$ with radius $r$ is defined to be the squared distance of $p$ from $q$, minus the squared radius; it is positive when $p$ is outside the disk (equalling the squared length of a tangent line segment from $p$ to the disk) and negative when $p$ is inside the disk. The \emph{power diagram} of a set of disks (Figure~\ref{fig:powerdiag}) is a partition of the plane into cells, at most one cell per disk, such that the points within a disk's cell have a smaller value of the power for that disk than for any other disk; it is known to have polygonal cells, and to be constructible in time $O(n\log n)$~\cite{Aur-SJC-87}.  The planar dual to a power diagram is a \emph{regular triangulation} having one vertex for each of the cells in the power diagram. Note that, unlike the Voronoi diagrams and Delaunay triangulations that they generalize, power diagrams and regular triangulations need not have a cell or a vertex for every input disk.

\begin{lemma}
\label{lem:power-center}
If $A$ is a disk in a collection $S$ of disks, $p$ is the center of $A$, and $B$ is the disk corresponding to the region of the power diagram that contains $p$, then either $A=B$ or $A$ and $B$ intersect.
\end{lemma}

\begin{proof}
The power of $p$ with respect to $A$ is negative. Thus, if $p$ is not in the cell of the power diagram corresponding to $A$, it must have an even more negative power with respect to the disk $B$ defining that cell, and therefore be inside $B$. Thus, the intersection of $A$ and $B$ contains the point $p$.
\end{proof}

\begin{lemma}
\label{lem:power-spans}
Suppose that disks $A$ and $B$ intersect and are part of a collection $S$ of disks. Then there exists a path in the regular triangulation dual to the power diagram of $S$, such that the starting vertex in the path corresponds either to $A$ or to a circle intersecting $A$, the ending vertex in the path corresponds either to $B$ or to a circle intersecting $B$, and each adjacent pair of vertices in the path is dual to a pair of intersecting disks.
\end{lemma}

\begin{proof}
Draw a curve from the center of $A$ to the center of $B$ that stays within $A\cup B$ and does not pass through any vertex of the power diagram. This curve passes through a sequence of cells of the power diagram corresponding to a path in the dual regular triangulation. It has negative power with respect to either $A$ or $B$ at each of its points, and therefore negative power with respect to each of the disks corresponding to the cells it passes through. Therefore, when it crosses a boundary from one cell to the next, it has negative power with respect to both corresponding disks, showing that the two disks have a nonempty intersection containing that crossing point.
\end{proof}

\begin{theorem}
Given a set of disks in the plane, we can construct a spanning forest of the intersection graph of the disks in time $O(n\log n)$.
\end{theorem}

\begin{proof}
Construct the power diagram and the dual regular triangulation, form the subgraph $G$ of the regular triangulation consisting of pairs of vertices corresponding to intersecting pairs of disks, and compute a spanning forest $F$ of $G$. For every disk $A$ not corresponding to a vertex in this spanning forest, let $p$ be the center of $A$, let $B$ be the disk whose power diagram cell contains $p$, create a new vertex representing $A$, and add an edge in $F$ connecting this new vertex to the vertex representing $B$.

By construction and by Lemma~\ref{lem:power-center}, every edge in the resulting forest represents a pair of disks that intersect, so it is a subforest of the intersection graph of the disks. By Lemma~\ref{lem:power-spans}, every intersection graph edge may be replaced by a path in $F$, so it is a spanning subforest of the intersection graph.
\end{proof}

As we now show, we cannot hope to find a similarly fast bound for connectivity of balls in higher dimensions, unless we improve on other more well-studied problems.

\begin{theorem}
\label{thm:con=mst}
The best possible randomized expected time bounds for connectivity of unit balls in $\R^d$ and Euclidean minimum spanning trees in $\R^d$ are within constant factors of each other.
\end{theorem}

\begin{proof}
In one direction, a spanning forest of the intersection graph of unit balls can be found as a subforest of the Euclidean minimum spanning tree of the ball centers.

In the other direction, suppose we have an algorithm that in time $T(n)$ can find the connected components of the intersection graph of a collection of $n$ unit balls.
Then we can test whether, in a set of $n$ red and blue points, there exists a red-blue pair at distance closer than two: compute the connected components  of balls centered at the points and test whether any component contains points of both colors.  A randomized reduction of Chan~\cite{Cha-DCG-99} transforms this decision algorithm into an optimization algorithm with the same complexity, for finding the closest red-blue pair, and it is known that this bichromatic closest pair problem and the Euclidean minimum spanning tree problem have asymptotically equal complexities~\cite{AgaEdeSch-DCG-91,Cal-PhD-95,KrzLev-ESA-97}.
\end{proof}

Recently, F{\"u}rer and Kasiviswanathan have showed that spanners for unit ball intersection graphs may be computed nearly as efficiently as spanning forests of these graphs~\cite{cs.CG/0605029}. Their results add spanner construction to the collection of problems with roughly equivalent algorithmic complexities given by Theorem~\ref{thm:con=mst}.

\section{Sweep for segments}

Next, we describe an efficient algorithm for our original motivating problem: bipartiteness of intersection graphs of line segments.  The generic algorithm for this case seems to involve complex ray shooting data structures and does not approach linear time.  Instead, we describe an alternative approach based on the plane sweep paradigm. We assume throughout that the arrangement is in \emph{general position}: no segment endpoint lies on another segment, no two segment endpoints or crossing points have the same $x$ coordinate, and no three segments cross at the same point. These conditions can be achieved, as is standard in computational geometry, by a symbolic perturbation of the data, at a constant cost in time per operation on the perturbed data, without changing the intersection graph of the input: in particular, in order to ensure that intersections occurring at segment endpoints are not eliminated by this perturbation, we slightly lengthen each segment before performing the other perturbations necessary to achieve general position. We refer the reader to Yap~\cite{Yap-SCG-88} and Edelsbrunner and M\"ucke~\cite{EdeMue-TOG-90} for details.

\begin{figure}[t]
\centering\includegraphics[height=2.5in]{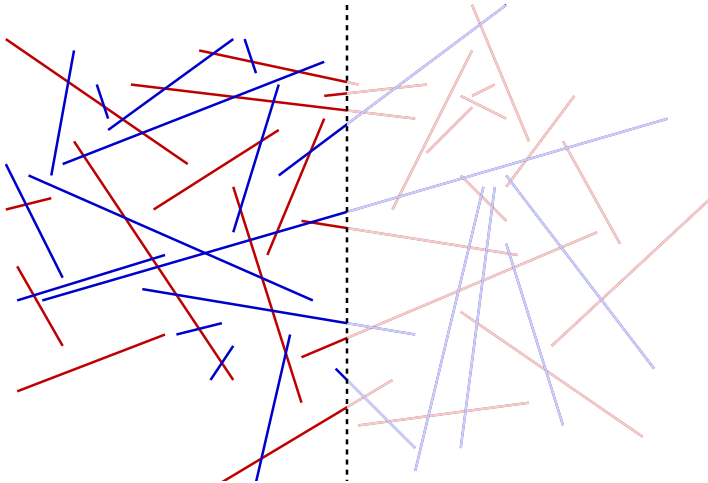}
\caption{Sweeping a vertical line (shown as dashed) from left to right across a bipartite arrangement of line segments. The truncated arrangement is the more strongly colored part of the plane to the left of the sweep line.}
\label{fig:truncated-arrangement}
\end{figure}

Our algorithm sweeps a vertical line left to right across the arrangement.
We define the \emph{truncated arrangement} to be the arrangement of line segments
formed by intersecting each input line segment with the halfplane to the left of the sweep line (Figure~\ref{fig:truncated-arrangement}).
As the algorithm progresses, we maintain various data structures, detailed below, to help it keep track of the connected components of the intersection graph of the truncated arrangement, and of a two-coloring of each component.
Also, as with most plane sweep algorithms, we use a priority queue to maintain a set of potential \emph{events} at which the other structures being maintained by the algorithm change combinatorially.  Each event is associated with a point in the plane, and the $x$ coordinate of this point is used as the event's priority, so that the priority queue ordering corresponds to the order in which the sweep line crosses the corresponding points.

\begin{figure}[t]
\centering\includegraphics[height=2.5in]{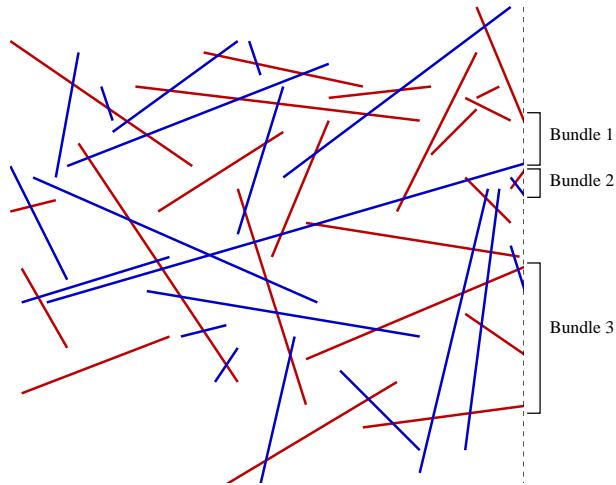}
\caption{Bundles of the truncated arrangement: sets of segments that cross the sweep line consecutively and are connected to each other in the truncated arrangement.}
\label{fig:bundles}
\end{figure}

\begin{figure}[t]
\centering\includegraphics[height=1.5in]{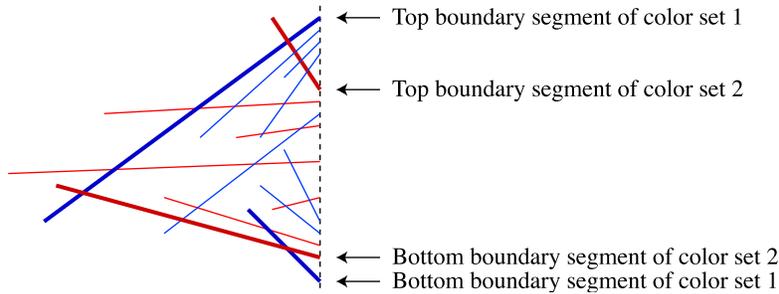}
\caption{The four boundary segments of a bundle.}
\label{fig:bundle-boundary}
\end{figure}

At any point in the algorithm, the sweep line will cross a subset of segments from the input arrangement, and these segments can be ordered by the $y$-coordinates of the crossing points.  This \emph{crossing sequence} changes combinatorially only when the sweep line crosses an endpoint of a segment or a crossing point of two segments.
We define a \emph{bundle} to be a maximal set of segments, appearing contiguously within the crossing sequence, and all belonging to the same connected component of the truncated arrangement (Figure~\ref{fig:bundles}).  Thus, the bundles form a partition of the crossing sequence into contiguous subsequences.  Note that a single component of the truncated arrangement may have several or no bundles associated with it. Within a bundle, we will be able to partition the segments into two different color sets, although we may not have fixed the choice of which color is assigned to which of these sets. At any point in the traversal, each bundle has up to four \emph{boundary segments} associated with it: the line segments in each of the two color sets that intersect the sweep line at the topmost and bottommost points (Figure~\ref{fig:bundle-boundary}

Our algorithm will maintain several data structures as it traverses the arrangement. The first, which we call the \emph{bundle tree}, is a balanced binary search tree with one tree node per bundle, representing the top-to-bottom order in which the bundles meet the sweep line. Each node in the bundle tree stores with it the boundary segments of the corresponding bundle.

\begin{lemma}
\label{lem:bundle-tree}
The bundle tree can be implemented so that, in logarithmic time per operation, it can handle the following operations:
\begin{itemize}
\item Split one bundle into two adjacent bundles, specifying the boundary segments of the two new bundles.
\item Merge two adjacent bundles into a single bundle.
\item Insert a new bundle between two existing bundles.
\item Remove a bundle.
\item Change the set of boundary segments associated with a bundle.
\item Search for a query point on the sweep line, and determine which bundle, if any, contains that point. If the point is not contained within a bundle, the result of the query should be the bundles closest to the query point above and below it.
\end{itemize}
\end{lemma}

\begin{proof}
Insertion and removal are standard binary search tree operations. Changing the set of boundary segments is simply a change of the data stored at each tree node and can be handled in constant time. Split operations can be handled as an insertion, followed by changing the boundary segments on the newly inserted node and the existing node associated with the two new bundles. Merge operations can be handled as a deletion, followed by changing the boundary segments on the remaining node by combining the boundary segment information from the two merged bundles. And a search for a query point can be handled by a binary search in the search tree, using the boundary segments of each bundle to determine which direction the search should follow in each of its steps.
\end{proof}

The second data structure we maintain is an \emph{event queue}: a priority queue of a set of \emph{events} at which the structures maintained by the algorithm need to change. We will process events of three types: the left endpoints of line segments, the right endpoints of segments, and the crossing points of pairs of segments. These are prioritized by their $x$-coordinates, with ties broken by the $y$-coordinate of the event point; we may assume without loss of generality by perturbing the input if necessary without changing its set of crossings, that no two events happen at the same event point. The event queue will store all segment endpoints to the right of the sweep line. In addition, it will store all crossing points between two line segments that are the boundary segments of adjacent bundles. However, crossings of other types will not be included in the event queue.

\begin{lemma}
\label{lem:event-queue}
The event queue described above, with the set of event points as described above, may be maintained under the operations described in Lemma~\ref{lem:bundle-tree}, and each sucessive event may be found and removed from the queue, in logarithmic time per operation.
\end{lemma}

\begin{proof}
We use a standard priority queue data structure such as a binary heap, with the priorities stated above in which a point is prioritized by its $x$ coordinate and then by its $y$ coordinate. Such a data structure can handle event insertions and deletions, and queries asking for the next event, in logarithmic time. To the information stored in the bundle tree we add a pointer from each bundle to the corresponding crossing point events in the event queue. When we merge two adjacent bundles, we remove from the event queue any crossing event both of the segments of which are from these two bundles. When we split a bundle into two, we add to the event queue any crossing event that can be formed by the boundary segments of the two new bundles. When we insert a new bundle, we remove from the event queue any crossing events involving its two neighbors, and add to the event queue any crossing events involving it with one of its two neighbors. And when we remove a bundle, we remove from the event queue any crossing events involving it and add to the event queue any crossing events that can be formed from the boundary segments of its two neighbors. In this way, each of the operations from Lemma~\ref{lem:bundle-tree} can be handled using a constant number of priority queue operations.
\end{proof}

The third data structure used by our algorithm is the \emph{parity forest}. This is a representation of a spanning forest of the truncated arrangement, with an arbitrarily chosen root for each tree of the forest. Each node in the parity forest represents a segment of the truncated arrangement. The \emph{parity} of a node is odd if the path from the node to the root of its tree has odd length, and even if the path has even length.

\begin{figure}[t]
\centering\includegraphics[width=4in]{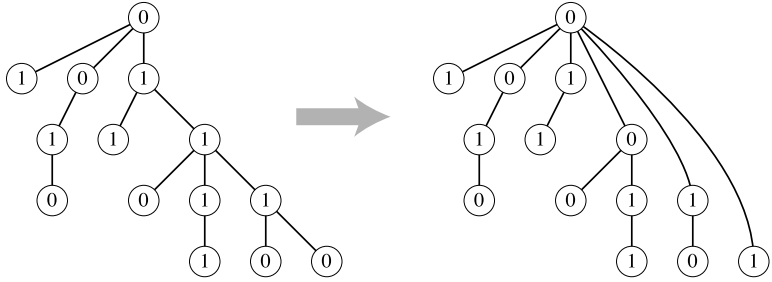}
\caption{Path compression in a parity forest. The parity bits stored at each node are adjusted so that the total parity of the path from each node to the root remains unchanged.}
\label{fig:path-compression}
\end{figure}

\begin{lemma}
\label{lem:parity-forest}
The parity forest may be maintained, subject to the following set of operations, in $O(\alpha(n))$ time per operation, where $\alpha$ denotes the inverse Ackermann function:
\begin{itemize}
\item Create a new node, forming a new tree in the forest.
\item Determine whether two nodes belong to the same trees or different trees.
\item Report the parity of a given node.
\item Add an edge between two nodes that belong to different trees, choosing arbitrarily a root for the new tree.
\end{itemize}
In addition, we may at any time list all edges in the parity forest, in linear time.
\end{lemma}

\begin{proof}
We modify the union-find data structure~\cite{Tar-JACM-75} to handle the desired operations. Recall that the union-find structure represents a family of disjoint sets as a forest, and identifies each set with the root of its tree in the forest. Each root node of the forest is labeled by the size of the tree descending from it. The union-find structure handles queries for the identity of a set containing a given element by following the path in the forest from that element to the root, and \emph{compresses} the path by reconnecting each element in it directly to the root.  It handles union operations between pairs of sets by adding an edge from the root of the smaller set to the root of the larger set and updating the sizes stored at the root of the combined tree.

To handle parity forest operations, we augment each node of a union find structure by a single parity bit. This bit is always zero for the root of a tree in the union find structure, but may be nonzero for other nodes. The parity of a node is determined as the sum (modulo 2) of the parity bits on the path from that node to the root of the tree, including the bit stored at the node itself. Determining whether two nodes belong to the same tree is handled by performing a find operation and then comparing the resulting two tree roots.
When a find operation compresses a path, the parity bits of the updated nodes are updated to equal the parities of the paths to those nodes, as depicted in Figure~\ref{fig:path-compression}. To report the parity of a node, perform a find operation and then examine its parity bit.

The most complex operation in this data structure is the operation of adding an edge between two nodes $x$ and $y$. Note that the union-find structure's forest need not have the same structure as the forest formed by these edge operations (although it will have the same components) and that $x$ and $y$ need not be the roots of their trees. We perform a union operation on the two trees containing $x$ and $y$, keeping the parities of nodes in the larger tree unchanged. However, we must update the parities of the nodes in the smaller tree in order to give $x$ and $y$ unequal parities. To do so, before performing the union operation, we look up the parity of $x$ and $y$ in their respective trees. If these two parities are unequal prior to the union, they will remain unequal after the union, because the parity bit stored at the root of each tree is zero and the paths after the union will differ from those before only in the addition of a root node. However, if before the union the parities of $x$ and $y$ are equal, we set the parity bit of the smaller tree's root to one before adding an edge in the union find structure connecting it to the larger tree's root. This will change the parities of all nodes in the smaller tree, preserving a correct partition of the nodes into two sets of opposite parity but forcing $x$ and $y$ to have different parities after the merge.

Each parity forest operation takes a constant number of union-find data structure operations; therefore, the time per operation is $O(\alpha(n))$. In order to report all edges efficiently, we also maintain as part of the data structure a list of all edges. When we add an edge to the parity forest we add it to this list, and when asked to report all edges in the forest we output this list.
\end{proof}

Our final data structure, the \emph{color tree}, represents the sorted order of segments crossing the sweep line among the line segments that are in the same color class of the same connected component of the truncated arrangement. In order to bound the complexity of the operations in this structure, we need some technical lemmas regarding the ordering of the segments belonging to different components in the truncated arrangement.

\begin{figure}[t]
\centering\includegraphics[height=1.5in]{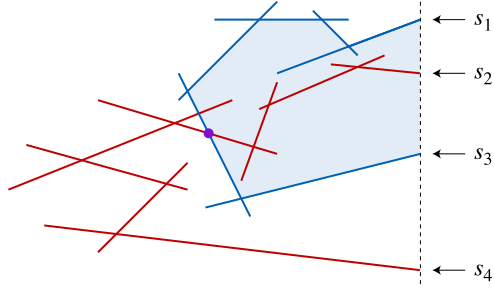}
\caption{Illustration for Lemma~\ref{lem:DS-seq}: if $s_1$ and $s_3$ belong to a single connected component of the truncated arrangement, the path connecting them (blue) together with a segment of the sweep line, forms a Jordan curve separating one region of the plane from another (light blue). If $s_2$ and $s_4$ are also connected by a path in the truncated arrangement (red), the two paths must cross at least once (the purple disk), showing that all four segments belong to a single component.}
\label{fig:DS-seq}
\end{figure}

\begin{lemma}
\label{lem:DS-seq}
Let $s_1$, $s_2$, $s_3$, and $s_4$ be four segments that cross the sweep line in the given top-to-bottom order. Then it is not possible that $s_1$ and $s_3$ belong to one component of the truncated arrangement, and that $s_2$ and $s_4$ belong to a different component of the truncated arrangement.
\end{lemma}

\begin{proof}
If $s_1$ and $s_3$ belong to a single component of the truncated arrangement, there must be a simple path $P_1$ within the segments of that component of the truncated arrangement that connects $s_1$ to $s_3$; $P_1$, together with the segment of the sweep line that lies between $s_1$ and $s_3$, forms a Jordan curve which by the Jordan curve theorem separates the points where $s_2$ and $s_4$ cross the sweep line. If $s_2$ and $s_4$ also belong to a single component of the truncated arrangement, there must be a second path $P_2$ within the segments of that component the truncated arrangement connecting them; but $P_1$ and $P_2$ would have to cross, so in this case the components containing the two paths could not be different from each other.
\end{proof}

The proof of this lemma is illustrated in Figure~\ref{fig:DS-seq}.
Lemma~\ref{lem:DS-seq} can be rephrased as saying that the sequence of component identities of the segments that cross the sweep line forms a Davenport-Schinzel sequence of order two~\cite{ShaAga-95}. As is standard for Davenport-Schinzel sequences of this order, this gives us a nesting structure among the components, which we express more formally in the following lemma.

\begin{figure}
\centering\includegraphics[height=2in]{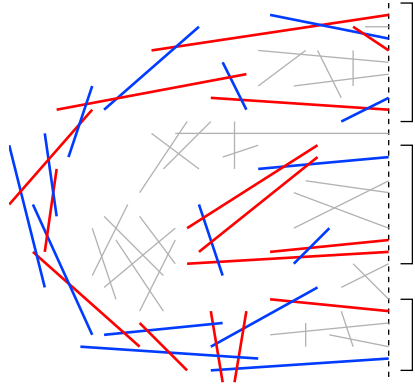}
\caption{Lemma~\ref{lem:color-nest}: two components of the truncated arrangement cross the sweep line in at most three contiguous subsequences of segments belonging to a single component.}
\label{fig:component-subsequences}
\end{figure}

\begin{lemma}
\label{lem:color-nest}
Let $C_1$ and $C_2$ be two connected components of the truncated arrangement, both crossing the sweep line, and let $s_i$ ($i=0, 1, \ldots$) be the top-to-bottom sequence of segments in $C_1\cup C_2$ that cross the sweep line. Then this sequence can be partitioned into at most three contiguous subsequences such that the segments within one of the subsequence belong to a single component.
\end{lemma}

\begin{proof}
Partition the sequence into maximal contiguous subsequences, such that within any of the subsequences each segment belongs to the same component. Then there cannot be more than three subsequences, for if there were four or more then choosing one segment each from the first four subsequences would result in a violation of Lemma~\ref{lem:DS-seq}.
\end{proof}

The three subsequences of this lemma are depicted in Figure~\ref{fig:component-subsequences}. Note that the subsequences of the lemma need not coincide with the bundles of the truncated arrangement, because they may be interrupted by segments from components other than $C_1$ and $C_2$; these segments from other components are shown in grey in the figure.

We define a sequence of segments $s_i$ ($i=0, 1, \ldots$) to be \emph{sequentially non-crossing} if each consecutive pair $s_i$, $s_{i+1}$ is disjoint. In any bipartite arrangement of segments, the sequence of segments of a single color class within a single component of a truncated arrangement must be sequentially non-crossing, because this is a weaker condition than the condition defining bipartiteness, that all segments within a single color class must be disjoint.

\begin{lemma}
\label{lem:color-tree}
We can define a data structure, the \emph{color tree}, that can maintain a set of sequentially non-crossing segments for a single color class of a single component of the truncated arrangement, subject to the following operations, in logarithmic amortized time per arrangement:
\begin{itemize}
\item Create a new color tree for a new line segment belonging to its own connected component of the truncated arrangement.
\item Remove a line segment from the color tree and verify that the remaining segments in the tree are sequentially non-crossing. If it is not sequentially non-crossing, report a crossing pair of segments.
\item Merge two components of the truncated arrangement, specified by a pair of crossing segments from the two components, combine their color trees, and verify that the segments represented by the merged color trees remain sequentially non-crossing. If it is not sequentially non-crossing, report a crossing pair of segments. We assume that, prior to this operation, the parity forest has not yet been updated to merge its representation of the same two components due to the same crossing, and update it as part of the merger process.
\end{itemize}
\end{lemma}

\begin{proof}
We represent a color tree using a binary search tree data structure that is capable of split and concatenate operations, such as a splay tree; the ordering for the binary search tree will be the top-to-bottom order in which the represented segments cross the sweep line. To keep track of the color trees for each component of the truncated arrangement, we will maintain two pointers from each root of the parity forest (representing a component of the truncated arrangement) to the color trees representing its two color classes. The first of the two pointers will be to the color tree with the same color as the root segment of the parity forest, and the second pointer will be to the other color tree for the same component. It is straightforward to maintain these pointers as the parity forest is updated according to Lemma~\ref{lem:parity-forest}.

Creating a new color tree is then simply a matter of creating a new binary search tree object with a single node, looking up the given segment in the parity forest, and pointing from the resulting parity forest root to the new color tree. To remove a segment from the color tree, simply check whether the two neighboring segments in the same tree cross each other; if they do, report that the result of the deletion would not be sequentially non-crossing, and otherwise delete the node corresponding to the given segment from the binary search tree.

To merge two components $C_1$ and $C_2$, we search for the given segments in the parity forest to identify the color trees associated with the merged components.  We then add an edge connecting the given segments to the parity forest, as described in Lemma~\ref{lem:parity-forest}.
By finding the topmost and bottommost segment within each color tree of each component, and comparing the positions of these segments in the top-to-bottom order as they cross the sweep line, we may determine (according to Lemma~\ref{lem:color-nest}) how to partition the segments of $C_1$ and $C_2$ into at most three contiguous subsequences, within each of which the segments belong to a single one of the given component. If there are three subsequences, rather than two, we perform split operations on the color trees for one of the two components so that the resulting binary search trees represent the contiguous orderings of the segments within a single color class within each of these three subsequences. Thus, we have at most six binary search trees associated with the split pieces of the merged components $C_1$ and $C_2$. We then concatenate these binary search trees into two trees, one for each color class of the newly merged component, using the parity forest to determine which pairs of binary search trees represent segments of the same color that should be concatenated. Finally, we update the pointers from the parity forest root representing the merged component, to point to the trees resulting from this concatenation.

Each operation takes a constant number of binary search tree and parity forest operations, so the amortized time per operation is logarithmic.
\end{proof}

We next go through a sequence of lemmas showing how to process each event stored in the event queue.

\begin{lemma}
\label{lem:event-left}
Suppose that the data structures described above correctly represent the components and two-coloring of the truncated arrangement just prior to the sweep line crossing the left endpoint of a line segment. Then we may update the data structures after that crossing so that they still correctly represent the components and two-coloring of the truncated arrangement, in logarithmic amortized time.
\end{lemma}

\begin{proof}
When the sweep line crosses the left endpoint of the segment, it cannot yet have swept across any crossing point involving that segment, so that segment must belong to its own new component of the truncated arrangement. By searching for the endpoint of the new segment in the bundle forest, we may determine whether it lies within an existing bundle or between two bundles; if it lies within an existing bundle, we must split that bundle in two parts above and below the new segment. To do so, we search for one of the bundle boundary segments in the parity forest to find the color tree representing the connected component of the bundle, and binary search for the new segment endpoint in the color trees for that component to find the new boundary segments of the resulting bundles after this split. We then split the bundle supplying these new boundary segments as arguments to the split operation. After this split has been performed, if necessary, we insert a new bundle for the new segment into the bundle tree. As described in Lemma~\ref{lem:event-queue}, we update the event queue to reflect these changes in the bundle tree. We create a new tree for the new segment in the parity forest, create a new color
tree for the new segment, and set a pointer from the parity forest tree's node to the color tree.

Thus, the left endpoint event can be handled using a constant number of operations in the bundle tree, parity forest, and color trees, so by the lemmas describing those data structures it takes logarithmic amortized time.
\end{proof}

\begin{lemma}
\label{lem:event-right}
Suppose that the data structures described above correctly represent the components and two-coloring of the truncated arrangement just prior to the sweep line crossing the right endpoint of a line segment. Then we may update the data structures after that crossing so that they still correctly represent the components and two-coloring of the truncated arrangement, in logarithmic amortized time. If this update causes the sequential non-crossing property of the color trees to fail, an odd cycle in the arrangement exists and may be reported in linear time.
\end{lemma}

\begin{proof}
We search the bundle tree to find the bundle containing the segment whose endpoint is being crossed. If that segment is the only boundary segment of its bundle, it is the only segment in that bundle; in this case, we delete that bundle from the bundle tree, check (by searching for the boundary segments in the parity forest) whether its two neighboring bundles belong to the same connected component of the truncated arrangement, and if so merge them. If, on the other hand, the segment is not the only remaining segment in its bundle, but is one of the boundary segments of the bundle, we must update the boundary segments of that bundle; to do so, we use the parity forest to find the color tree containing the given segment, find its neighboring segments in the color tree, and check those neighboring segments against the boundary segments of the neighboring bundles in the bundle tree to determine which of them belongs to the same bundle and can be used as a replacement. Then, in either case, after the bundle tree has been updated, we update the event queue as described in Lemma~\ref{lem:event-queue} to reflect these changes.

Finally, we use the parity forest to find the color tree for the segment whose endpoint is being crossed, remove that segment from the color tree, and check that the color tree remains sequentially non-crossing. If a violation of the sequential non-crossing property is found, it means that two segments belonging to the same color class of the same component cross; in this case the intersection graph edge between these two crossing segments, together with the path connecting them in the parity forest, forms an odd cycle in the intersection graph.
\end{proof}

\begin{lemma}
\label{lem:event-cross}
Suppose that the data structures described above correctly represent the components and two-coloring of the truncated arrangement just prior to the sweep line passing across the crossing point between two boundary segments $s_i$ and $s_j$ of two adjacent bundles. Then we may update the data structures after that crossing so that they still correctly represent the components and two-coloring of the truncated arrangement, in logarithmic amortized time. If this update causes the sequential non-crossing property of the color trees to fail, an odd cycle in the arrangement exists and may be reported in linear time.
\end{lemma}

\begin{proof}
The two bundles must represent different components of the truncated arrangement, by the definition of a bundle. We merge the color trees for these components (as part of the merge process also merging their trees in the parity forest) as described in Lemma~\ref{lem:color-tree}. If a violation of the sequential non-crossing property is found, it means that two segments belonging to the same color class of the same component cross; in this case the intersection graph edge between these two crossing segments, together with the path connecting them in the parity forest, forms an odd cycle in the intersection graph. Let $N$ denote the set of segments that, after this color tree merger, have a neighbor in the color tree that formerly belonged to a different component; $N$ has constant size and may be constructed in constant additional time as we perform the color tree merge.

We then also merge pairs of adjacent bundles in the bundle tree that belong to the two merged components. One of these pairs of bundles must be the one containing segments $s_i$ and $s_j$, and by Lemma~\ref{lem:color-nest} there may be at most one other such pair; if this second pair of adjacent bundles to be merged exists, one of the boundary segments in each bundle must belong to $N$. Thus, these pairs of bundles to be merged may be found by using the bundle tree to search for the segments in $N$, determining which bundles each segment in $N$ may be a boundary segment of, and then testing whether the neighboring bundles in the bundle tree belong to the same merged component. Thus, there are a constant number of pairs of bundles to merge, each of which may be found using a constant number of bundle tree search operations. After the bundle tree has been updated, we update the event queue as described in Lemma~\ref{lem:event-queue} to reflect these changes.
\end{proof}

The overall algorithm we follow consists of handling events one by one from the event queue, as described in the lemmas above; it is outlined as pseudocode in Table~\ref{tbl:segsweep}.

\begin{table}
{\small
\begin{lstlisting}
initialize event queue Q to contain all segment endpoints
initialize parity forest P, bundle tree B, and color tree C to be empty
while Q is nonempty:
    find and remove leftmost event point e from Q
    if e is the left endpoint of a line segment s:
        search B for the bundle b containing e
        if e is between the upper and lower boundary segments of b:
            split b into two bundles in B
        insert s as a new bundle in B
        update Q to reflect the changed sequence of adjacent bundles in B
        create and link new nodes in P and C for s
    elif e is the right endpoint of a line segment s:
        search B for the bundle b containing e
        search P for the color tree c containing s
        if s is the only segment of b:
            remove b from B
            update Q to reflect the changed sequence of adjacent bundles in B
        elif s is a boundary segment of b:
            use c to find a replacement boundary segment for s
            change the boundary segments of b
            update Q to reflect the changed boundary segments of b
        remove s from c
        if the updated color tree is not sequentially noncrossing:
            let a and b be two edges that cross each other
            use P to find an even path p in the intersection graph from a to b
            return the odd cycle formed by p and the edge ab
    elif e is the crossing between two segments s and t:
        use P to find the color trees containing s and t
        merge the two color trees and add edge st to P
        if the merged color tree is not sequentially noncrossing:
            let a and b be two edges that cross each other
            use P to find an even path p in the intersection graph from a to b
            return the odd cycle formed by p and the edge ab
        let N be the set of segments at which the color trees were merged
        for each segment n in N:
            search B for the bundle b containing n
            use P to find the components of b and its neighbors in B
            if b and a neighbor belong to the same components as s and t:
                merge b and its neighbor in B
        update Q to reflect the changed sequence of adjacent bundles in B
initialize a coloring X
for each input segment s:
    if parity of s in P is even:
        X[s] = red
    else:
        X[s] = blue
return the bipartite coloring X
\end{lstlisting}}
\caption{Pseudocode of line segment bipartiteness testing algorithm.}
\label{tbl:segsweep}
\end{table}

\begin{lemma}
\label{lem:correct-events}
The information about the truncated arrangement represented by the data structures described above does not change except at the events of the type described above, unless one of the events leads to the detection of an odd cycle in the arrangement's intersection graph.
\end{lemma}

\begin{proof}
Because the information represents the combinatorial properties of the truncated arrangement, it only changes when the sweep line passes over a segment endpoint or the crossing between two segments. Crossings between segments representing different components of the truncated arrangement must occur between boundary segments of bundles, and are represented as events. Crossings between segments within a single component that represent the same color must lead, just prior to the sweep line passing over the crossing, to a situation in which the color tree for that color of that component is not sequentially non-crossing, which would have already led to the detection of an odd cycle.

The only remaining crossing points are those between oppositely colored segments of the same component of the truncated arrangement. However, such crossings do not lead to a change in the bundles, bundle boundary segments, color trees, or sequential non-crossing property of the color trees.
\end{proof}

\begin{lemma}
\label{lem:few-events}
In an arrangement of $n$ segments, at most $3n-1$ events can be processed.
\end{lemma}

\begin{proof}
There are exactly $2n$ events arising from the endpoints of segments, so it remains to count the events corresponding to crossings between boundary segments of adjacent bundles. Let $\Phi$ denote the number of components of the truncated arrangement plus the number of left endpoints of segments that have not yet been crossed by the sweep line; $\Phi$ is initially $n$, is positive after all events are processed, is reduced by one when processing a crossing event because the event causes two components to merge, and is otherwise unchanged. Therefore, the total number of times it can be reduced, and the total number of crossing events, is at most $n-1$.
\end{proof}

\begin{theorem}
We can test the bipartiteness of an intersection graph of $n$ line segments in the plane, and return either a two-coloring of the intersection graph or an odd cycle, in time $O(n\log n)$.
\end{theorem}

\begin{proof}
We initialize the data structures to represent an empty structure and then process the events from the event queue as described in Lemma~\ref{lem:event-left}, Lemma~\ref{lem:event-cross}, and Lemma~\ref{lem:event-right}. Each event takes logarithmic amortized time to process, and by Lemma~\ref{lem:few-events} there are $O(n)$ events, so the total time is $O(n\log n)$. By Lemma~\ref{lem:correct-events} the structure of the truncated arrangement does not change except at the processed events, so it is correctly maintained by the data structures throughout the algorithm. If any step of the algorithm leads to an odd cycle being detected, the algorithm is terminated and the cycle reported in an additional $O(n)$ time. Otherwise, a two-coloring may be found by determining the parity of each segment in the parity forest.
\end{proof}

When the segments are found to be bipartite, the algorithm correctly identifies the connected components of their intersection graph, and constructs a spanning forest of this graph.  However, it is not able to perform the same connectivity analysis for nonbipartite segment arrangements.  It may be instructive to attempt to modify our techniques to handle nonbipartite arrangements, and observe in what way they run into difficulties: we may define the truncated arrangement and its bundles as before, and define the boundary segments of a bundle as the topmost and bottommost segments of the bundle in the order that they cross the sweep line. As before, the components of the truncated arrangement change only when the sweep line passes over a crossing between the bottom boundary segment of one bundle and the top boundary segment of another. However, in bipartite arrangements, the boundary segments of a bundle change only at segment endpoints, while in nonbipartite arrangements, the boundary segments of a bundle may change due to crossings within the bundle. Therefore, we cannot avoid handling crossings within a single bundle as we did in the bipartite case.

We observe that Hopcroft's problem, testing the existence of a point-line incidence in a set of $n$ points and $n$ lines, can easily be reduced to the problem of finding the connected components of a line segment intersection graph: simply view each point as a line segment with length zero (or a sufficiently small positive length), view
each line as a sufficiently long line segment (having endpoints outside the convex hull of the points), and test whether any point belongs to a nontrivial component of the intersection graph.  Hopcroft's problem has known $\Omega(n^{4/3})$ lower bounds, in natural and general models of geometric computation~\cite{Eri-DCG-96}.
Therefore, it seems unlikely that an algorithm with a near-linear time bound similar to that for bipartiteness can exist for the connectivity problem.

\section{Other planar objects}

The same approach works as well for more general classes of objects in the plane.

\begin{theorem}
We can test the bipartiteness of an intersection graph of a collection of simple polygons in the plane, having a total of $n$ sides, and return either a two-coloring of the intersection graph or an odd cycle, in time $O(n\log n)$.
\end{theorem}

\begin{proof}
We use a similar plane sweep approach and data structures, with some modifications.  First, the truncated arrangement now refers to the set of polygons formed by intersecting the input polygons with the halfplane left of the sweep line.  A single polygon may be represented as multiple polygons in the truncated arrangement, allowing us to preserve the nesting property used in the algorithm.  Second, the crossing sequence, and therefore also the color trees, may include the same polygon more than once.  For each appearance of a polygon in a color tree, we maintain the identities of the two polygon edges crossed by the sweep line, so that we can still perform binary searches in the color trees in constant time per step.
Third, we have a larger set of possible events.  Each polygon vertex causes an event, and these vertices may be classified according to how many incident edges go leftwards or rightwards from the vertex, and (in the case that both go leftwards or both go rightwards) whether the vertex is convex or reflex.  There are five possibilities:

\begin{figure}[t]
\centering\includegraphics[width=4in]{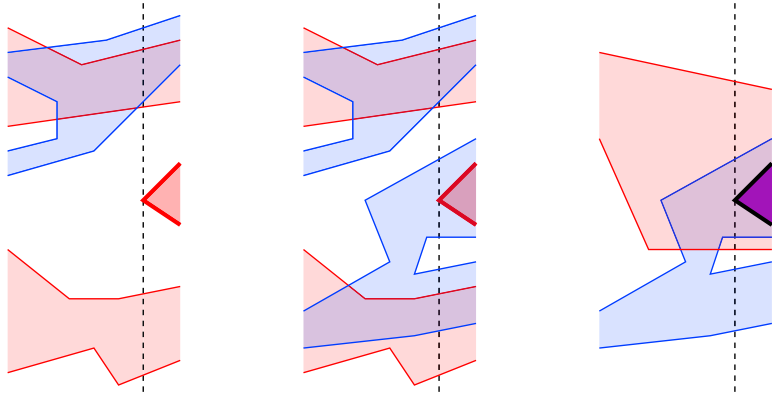}
\caption{Three subcases for a convex vertex with two edges extending rightwards: not part of any polygon and forming a new bundle (left), overlapping a single polygon and becoming part of an existing bundle (center), and overlapping two polygons and forming a triangle in the intersection graph (right).}
\label{fig:polygon-left}
\end{figure}

\begin{itemize}
\item A convex vertex with two rightward edges corresponds to the left endpoint of a line segment in the line segment algorithm. There are several different subcases that can arise for this type of vertex, depicted in Figure~\ref{fig:polygon-left}. If the vertex is not contained in any previous polygon of the truncated arrangement, this event is handled similarly to a left endpoint, by adding a new bundle to the bundle tree and splitting an existing bundle if necessary. However, it is possible that the vertex occurs within the interior of another polygon; we can detect this situation by searching for the vertex in the bundle tree and then in the appropriate color trees. In this case add the newly formed truncated polygon to an existing component and bundle of the truncated arrangement instead of creating a new component for it.
If the vertex occurs within two polygons of opposite colors in the same component,
its polygon and the two others form a triangle in the intersection graph, and we halt immediately.
\item A reflex vertex with two rightward edges (Figure~\ref{fig:polygon-cases}, left) corresponds to the split of one item into two items in the crossing sequence, and we split one node in two in the appropriate color tree. If, prior to sweeping across this vertex, its polygon was one of the boundary items of its bundle, the bundle should be updated to reflect the change to the crossing sequence, but the other data structures do not need to change.
\item A vertex with one leftward and one rightward edge (Figure~\ref{fig:polygon-cases}, center left) does not change the bundle tree or the crossing sequence, but it does change the identity of the edge crossed by the sweep line, and we make the appropriate change to the corresponding color tree node. This change may also result in modifications to the set of crossing events in the event queue, if the vertex's polygon was a boundary item of its bundle.
\item A reflex vertex with two leftward edges (Figure~\ref{fig:polygon-cases}, center right) corresponds to the merger of two polygons in the truncated arrangement into a single polygon.  We merge the two corresponding nodes in the appropriate  color tree.  If the two polygons previously belonged to different connected components of the truncated arrangement, we merge those components into a single one similarly to our treatment of a crossing event in our line segment algorithm.
\item A convex vertex with two leftward edges (Figure~\ref{fig:polygon-cases}, right) corresponds to the right endpoint of a line segment in the line segment algorithm, and is handled similarly to that case, by removing the item from the color tree and, if it was the last item in its bundle, removing the bundle from the bundle tree.
\end{itemize}

\begin{figure}[t]
\centering\includegraphics[width=4in]{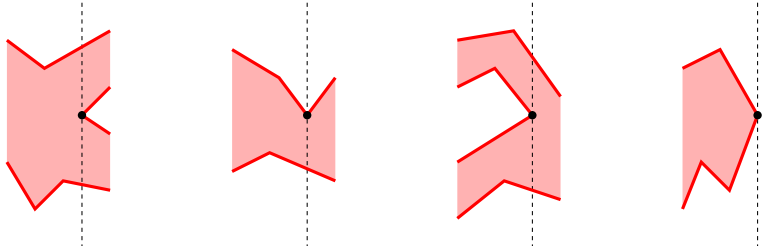}
\caption{Additional cases for the plane sweep algorithm for bipartiteness testing of polygon intersection graphs, from left to right: reflex vertex with two rightward edges, vertex with leftward and rightward edge, reflex vertex with two leftward edges, convex vertex with two leftward edges.}
\label{fig:polygon-cases}
\end{figure}

As before, there are $O(n)$ events processed, each of which involves a constant number of data structure operations, so the total time is $O(n\log n)$.
\end{proof}

\section{More than two colors}

Bipartiteness (two-coloring) of geometric intersection graphs can obviously be tested in polynomial time, and we have described efficient algorithms for several important cases of intersection graphs including the line segment intersection graphs arising from the problem of computing the geometric thickness of a drawing.  What about coloring with more than two colors?  For graphs, this is NP-complete, so it is not surprising that the same is true for geometric intersection graphs.
We outline a proof for planar unit disks and for line segments;
the result for unit disks was previously known~\cite{ClaColJoh-DM-90,GraStuWei-Algo-98} but for completeness we repeat it here.

\begin{figure}[t]
\centering\includegraphics[width=4.5in]{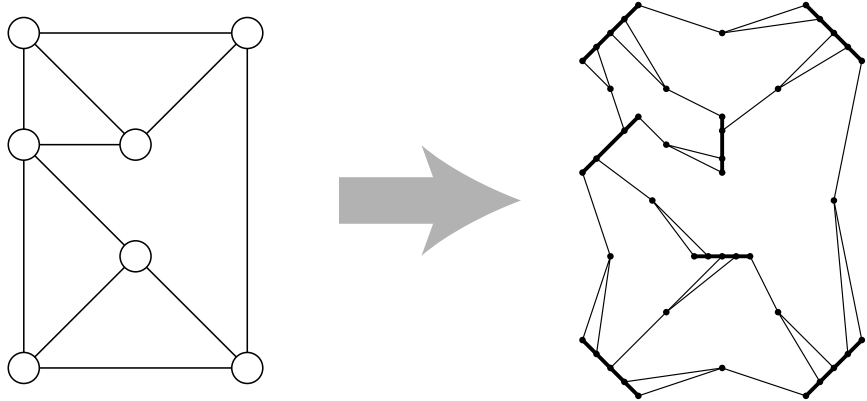}
\caption{Reduction for NP-completeness of 3-coloring segment intersection graphs. A grid-embedded planar graph (left) is transformed into a set of line segments (right) in such a way that 3-colorability is preserved. The thick line segments at right are the segments $s_v$ representing vertices, while the thin line segments are the segments $p_{vw}$, $q_{vw}$, and $r_{vw}$ representing edges; the small dark disks lie on the segments $s_v$ and at the midpoints of edges $vw$, and are points where two or three of the line segments intersect.}
\label{fig:segredux}
\end{figure}

\begin{theorem}
It is NP-complete, given a set of line segments in the plane, to determine whether the intersection graph of the segments is 3-colorable.
\end{theorem}

\begin{proof}
The problem is clearly in NP, as we may test whether a given 3-coloring of the segments is valid in polynomial time by constructing the intersection graph and testing each of its edges. To prove NP-completeness, we find a polynomial-time many-one reduction from a known NP-complete problem, 3-colorability of planar graphs~\cite{GarJohSto-TCS-76}. That is, given a planar graph $G$, we wish to construct in polynomial time a set $S$ of line segments that is 3-colorable if and only if $G$ is 3-colorable.

To construct $S$ from $G$, begin by finding a straight-line embedding of the graph on a polynomial size grid~\cite{Sch-SODA-90}. For each vertex $v$ in $G$, form a line segment $s_v$ centered on the point where $v$ is placed. We choose the angle of this line segment so that it does not lie on any line from $v$ to one of its neighbors. In addition, we choose the lengths of all line segments $s_v$ to be short enough so that, if $vw$ is an edge of $G$ with midpoint $x$ in the embedding, the triangle bounded by segment $s_v$ and point $x$ does not contain any other edge midpoint and is not crossed by any other segment $s_w$.

Next, for each edge $vw$ in $G$, form three line segments $p_{vw}$, $q_{vw}$, and $r_{vw}$, each having one endpoint on the midpoint of $vw$ in the embedding of $G$, the other endpoint on one of $s_v$ or $s_w$, with at least one endpoint on each of $s_v$ and $s_w$. Thus, $p_{vw}$, $q_{vw}$, and $r_{vw}$ are mutually intersecting, and each intersects one of $s_v$ and $s_w$; by choosing the placement of the endpoints on $s_v$ and $s_w$ in a way that is consistent with the cyclic order of the edges around $v$ and $w$ in the embedding, this construction can be performed in such a way that no intersections between segments exist other than the ones of these types. An example of this construction is shown in Figure~\ref{fig:segredux}.

We now claim that the set $S$ of segments so constructed is 3-colorable if and only if $G$ is 3-colorable. In one direction, suppose $G$ is 3-colorable. Then assign each segment $s_v$ the color of the corresponding vertex $v$. Among the three segments $p_{vw}$, $q_{vw}$, and $r_{vw}$, choose one that is not incident to $s_v$ and assign it the same color as $v$, and choose one that is not incident to $s_w$ and assign it the same color as $w$; assign the remaining segment the third color that is not used by $v$ and $w$. In this way, all incident pairs of segments are assigned distinct colors, so the result is a valid 3-coloring of $S$. In the other direction, suppose $S$ is 3-colorable. Then, for any edge $vw$ of $G$, segments $s_v$ and $s_w$ must be assigned distinct colors, for if they were both the same color it would not be possible to properly color $p_{vw}$, $q_{vw}$, and $r_{vw}$. Thus, assigning each vertex $v$ the color assigned to $s_v$ results in a proper coloring of $G$.

We have found a polynomial time construction for a set $S$ of line segments that is 3-colorable if and only if the given planar graph $G$ is 3-colorable; that is, we have a polynomial time many-one reduction from planar graph 3-coloring to segment intersection graph 3-coloring, showing that segment intersection graph 3-coloring is NP-hard. Since it is also in NP, it is NP-complete.
\end{proof}

Although our reduction produces sets of segments that intersect at the segment endpoints, it is straightforward to instead produce sets of segments that cross properly whenever they intersect, by slightly lengthening each segment of this construction.

\begin{figure*}[t]
\centering
\includegraphics[width=5.5in]{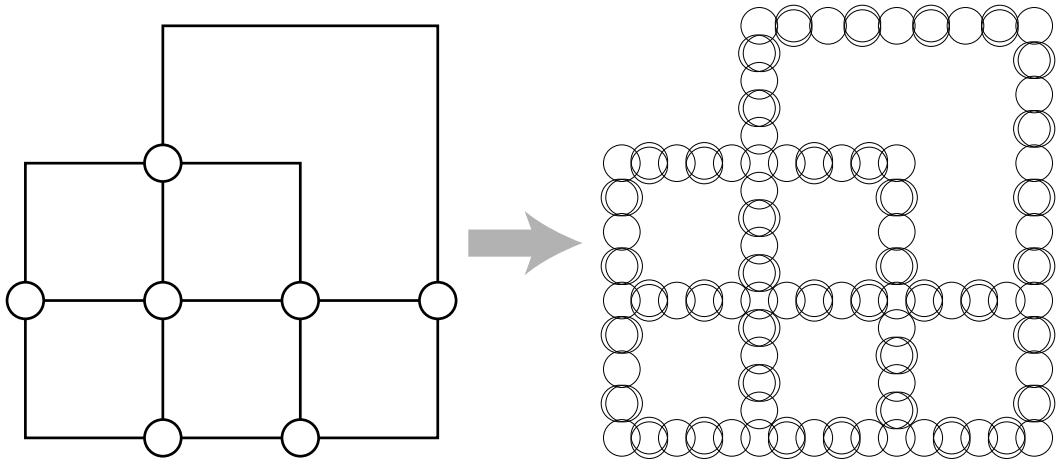}
\caption{NP-completeness reductions from degree-four planar graph 3-coloring to unit disk 3-coloring.}
\label{fig:udredux}
\end{figure*}

\begin{theorem}
It is NP-complete, given a set of unit disks in the plane, to determine whether the intersection graph of the disks is 3-colorable.
\end{theorem}

\begin{proof}
The problem is clearly in NP, as we may test whether a given 3-coloring of the disks is valid in polynomial time by constructing the intersection graph and testing each of its edges. To prove NP-completeness, we find a polynomial-time many-one reduction from a known NP-complete problem, 3-colorability of planar graphs with degree at most four~\cite{GarJohSto-TCS-76}. That is, given a planar graph $G$, with at most four edges per vertex, we wish to construct in polynomial time a set $S$ of unit disks that is 3-colorable if and only if $G$ is 3-colorable.

We begin our construction of $S$ by finding an orthogonal drawing of the graph on a grid, so that its edges are drawn with bends along grid edges~\cite{TamTol-A4C-89,DiBEadTam-99}. We expand the grid so that the vertices and corners of it are spaced exactly 7.5 units apart. We place a unit disk on each vertex or corner of the drawing, Along each grid edge that is covered by an edge in the graph drawing we place either four unit disks, equally spaced with centers 3/2 of a unit apart, or three unit disks, equally spaced with centers 15/8 of a unit apart; we choose four or three disks per grid edge in such a way that the grid path representing any edge in $G$ has an even number of disks placed along it.
Finally, along the grid path representing each edge, we replace every other disk by two disks with centers slightly perturbed from those of the replaced disk, so that they intersect each other, all disks that the replaced disk intersected, and no other disks. This construction is shown in Figure~\ref{fig:udredux}.

We now claim that $G$ is 3-colorable if and only if the resulting set $S$ of disks is 3-colorable. In one direction, suppose we have a 3-coloring of $G$. We may assign the disks corresponding to the vertices of $G$ the same colors; along every edge, the alternating sequence of pairs of disks and single disks has a unique coloring determined by the colors of the endpoints of the edge. Therefore, $S$ may be 3-colored. In the other direction, suppose that $S$ is 3-colored. The alternating sequence of pairs of disks and single disks along any edge of $G$ cannot be colored unless the disks at the two endpoints of the edge are colored differently; therefore, the colors assigned to the disks placed at the vertices of $G$ form a proper 3-coloring of $G$.

We have found a polynomial time construction for a set $S$ of unit disks that is 3-colorable if and only if the given planar graph $G$ is 3-colorable; that is, we have a polynomial time many-one reduction from planar graph 3-coloring to unit disk intersection graph 3-coloring, showing that unit disk intersection graph 3-coloring is NP-hard. Since it is also in NP, it is NP-complete.
\end{proof}

\section{Conclusions}

We have investigated several algorithms for testing bipartiteness of intersection graphs, one for arbitrary geometric objects, based only on the existence of a decremental intersection detection data structure, and two faster ones for fat objects (balls in $\R^d$) and planar objects (line segments or simple polygons).  We have also investigated the relation between bipartiteness testing and connectivity, and provided evidence that, for some types of geometric object, connected component analysis is strictly more difficult than bipartiteness testing.

There are several potential directions for future research.  Can the generic bipartiteness algorithm be improved upon for objects that are not fat and nonplanar, such as rectangular boxes or triangles in $\R^3$?  Might the chromatic number of a geometric intersection graph be easier to approximate than it is for general graphs?  For general graphs, bipartiteness can be maintained efficiently in a dynamic setting~\cite{EppGalIta-JACM-97,HolLicTho-JACM-01}; can bipartiteness of intersection graphs be maintained efficiently dynamically or kinetically?  Finally, for which other useful graph properties is it possible to provide similar speedups in the geometric setting over a naive approach that constructs the entire intersection graph?

\section*{Acknowledgements}

This work was supported in part by NSF grant
CCR-9912338. A preliminary version of this paper was presented at the 15th ACM-SIAM Symposium on Discrete Algorithms (SODA 2004). All figures are by the author, copyright of the author, and used by permission.

\raggedright
\bibliographystyle{abuser}
\bibliography{tbgig}
 \end{document}